\documentclass[twocolumn]{aastex701}

\usepackage{amsmath}

\begin{document}

\title{Blobs and Blurs: A Citizen Science-Identified Catalog of Diffuse Galaxies in the Fornax Cluster}

\author[0009-0005-9612-4722]{Nicolas Mazziotti}
\affiliation{Steward Observatory, University of Arizona, 933 North Cherry Avenue, Rm. N204, Tucson, AZ 85721-0065, USA}
\affiliation{Department of Astronomy, University of Illinois at Urbana-Champaign, Urbana, IL 61801, USA}
\email{nm78@illinois.edu}

\author[0000-0002-5434-4904]{Michael G. Jones}
\affiliation{Steward Observatory, University of Arizona, 933 North Cherry Avenue, Rm. N204, Tucson, AZ 85721-0065, USA}
\affiliation{IPAC, Mail Code 100-22, Caltech, 1200 E. California Blvd., Pasadena, CA 91125, USA}
\email{mgjones@ipac.caltech.edu}

\author[0000-0002-7013-4392]{Donghyeon J. Khim}
\affiliation{Steward Observatory, University of Arizona, 933 North Cherry Avenue, Rm. N204, Tucson, AZ 85721-0065, USA}
\email{galaxydiver@arizona.edu}

\author[0000-0003-4102-380X]{David J. Sand}
\affiliation{Steward Observatory, University of Arizona, 933 North Cherry Avenue, Rm. N204, Tucson, AZ 85721-0065, USA}
\email{dsand@arizona.edu}

\author[0000-0001-8354-7279]{Paul Bennet}
\affiliation{Space Telescope Science Institute, 3700 San Martin Drive, Baltimore, MD 21218, USA}
\email{pbennet@stsci.edu}
 
\begin{abstract}
We present a catalog of 643 diffuse galaxies identified through a citizen science search of the Fornax cluster, of which we estimate 21.8\% are nucleated (139/637; 6 inconclusive). This marks the first crowd-sourced effort to construct a cluster-scale census of diffuse galaxies. These objects were visually identified using a combination of the Fornax Deep Survey and Dark Energy Camera Legacy Survey imaging across 26 deg$^2$. Over 1,400 volunteers cataloged the candidates within this sky area at a rate of 1.15~days/deg$^2$. Our catalog is highly complete relative to existing dwarf catalogs of Fornax ($> 80\%$ of objects recovered) down to an effective radius $r_{\mathrm{eff}} = 5\arcsec$, the minimum size we suggested volunteers classify, and to an effective \textit{r}-band surface brightness as faint as $\langle \mu_r \rangle \simeq26$ mag arcsec$^{-2}$. We detect 97 candidates that existing automated searches of Fornax did not find and three candidates not found by any prior search, automated or visual. The stellar mass distribution of our sample is consistent with similar dwarf studies of Fornax, with the nucleated fraction peaking at 80\% for a host galaxy mass of $\sim$10$^{8.5}M_{\odot}$. The efficiency and completeness of our catalog thus establishes citizen science as a valuable tool for mapping diffuse galaxy populations in future sky surveys, such as the Legacy Survey of Space and Time.
\end{abstract}

\keywords{\href{http://astrothesaurus.org/uat/416}{Dwarf galaxies (416)}, \href{http://astrothesaurus.org/uat/1130}{Nucleated dwarf galaxies (1130)}, \href{http://astrothesaurus.org/uat/940}{Low surface brightness galaxies (940)}, \href{http://astrothesaurus.org/uat/584}{Galaxy clusters (584)}, \href{http://astrothesaurus.org/uat/1464}{Sky surveys (1464)}}

\section{Introduction} 
\label{sec:intro}
Dwarf galaxies are the most abundant galaxies in the universe \citep{Ferguson_1994}. They exist in all environments, with the highest concentrations residing in dense clusters such as the Virgo cluster \citep{Reaves_1983}. Dwarf galaxies that are more than a few Mpc away cannot readily be resolved into individual stars with current ground-based telescopes, and therefore have a diffuse, blur-like appearance where the light of all the stars blends together to form a smooth distribution. The radial profile of this light distribution generally follows a near-exponential form, which can be characterized by an effective radius ($r_\mathrm{eff}$) and central surface brightness ($\mu_0$). This class of galaxies are referred to as diffuse galaxies (DGs). A significant fraction of DGs contain a highly concentrated stellar population at their center known as a nuclear star cluster (NSC), thus forming distinct subclasses of nucleated and non-nucleated DGs. The most spatially extended diffuse dwarfs are known as ultra-diffuse galaxies (UDGs; \citealp{VanDokkum_2015}), defined by $r_\mathrm{eff} \ge$ 1.5 kpc and $\mu_{g,0} \ge$ 24 mag arcsec$^{-2}$. The properties of dwarf galaxies are strongly influenced by the local environment (e.g. structure and star formation history), making DGs good probes of astrophysical processes that can drive the evolution of satellite galaxies and powerful tests for $\Lambda$CDM cosmology (\citealp{Bullock_2017}; \citealp{Sales_2022}).

The Fornax cluster is the second nearest galaxy cluster at 19 Mpc \citep{Blakeslee_2009} and the closest high-density region in the southern sky. Older galaxy populations in the Fornax core region indicate that the cluster is dynamically evolved (\citealp{Ferguson_1989}; \citealp{Drinkwater_2001}), yet the cluster possesses a lopsided X-ray distribution that suggests the intracluster gas is not completely virialized \citep{Paolillo_2002}. With a virial mass $M_{vir} = 7 \times 10^{13} M_{\odot}$ and radius $R_{\mathrm{vir}} = 0.7$ Mpc \citep{Drinkwater_2001}, Fornax is less massive and more compact than larger clusters like Virgo ($M_{vir} = 6.3 \times 10^{14} M_{\odot}$, $R_{\mathrm{vir}} = 1.7$ Mpc) (\citealp{Mei_2007}; \citealp{Kashibadze_2020}) and Coma ($M_{vir} \simeq 10^{15} M_{\odot}$ and $R_{\mathrm{vir}} \simeq 2.3$ Mpc) (\citealp{Chernin2013}; \citealp{Keshet2017}). The richness of galaxies within a relatively small region of sky and close proximity to us makes Fornax a desirable target for statistical analyses of DG populations.

Complete catalogs of diffuse galaxies are challenging to produce because of their low surface brightness. In optical images, both visual and automated detection methods struggle to identify exceptionally faint cases and those obscured by bright objects, namely foreground stars. Improvements in imaging surveys of the nearby universe over the past decade have been vital for increasing completeness. The new generation of deep surveys of the Fornax and Virgo Clusters, such as the Fornax Deep Survey (FDS; \citealp{Peletier_2020}) and the Next Generation Virgo Survey (NGVS; \citealp{Ferrarese_2012}), have revealed galaxies with \textit{g}-band surface brightnesses as faint as 28.4 mag arcsec$^{-2}$ ($1\sigma$ noise averaged over 1 arcsec$^2$, \citealp{Peletier_2020}) and 29 mag arcsec$^{-2}$ ($2\sigma$ above mean sky level, \citealp{Ferrarese_2020}), respectively. Recent all-sky surveys like the Dark Energy Camera Legacy Survey (DECaLS/Legacy Survey; \citealp{Dey_2019}) and the forthcoming Legacy Survey of Space and Time (LSST; \citealp{LSST_2019}) allow for DGs to be detected across all types of cosmological environments, advancing the completeness of DGs universally (e.g., \citealp{Zaritsky_2019}). 

Large scale deep surveys have subsequently led to the development of automated detection algorithms specialized in finding DGs. \cite{Venhola_2018} established a detection pipeline built on Source Extractor \citep{Bertin_1996} to find dwarf galaxies in the Fornax Deep Survey using background subtracted \textit{gri} composite images, achieving a high level of completeness, and finding 564 dwarfs in total. \cite{Venhola_2022} extended this catalog to low surface brightness galaxies (LSBs) in Fornax via a max-tree based object detection algorithm, finding an additional 256 dwarfs with a mean \textit{r'}-band effective surface brightness $\gtrsim$ 23 mag arcsec$^{-2}$. A combination of a traditional extended source detection algorithm and a convolutional neural network (CNN) was utilized by Systematically Measuring Ultra-Diffuse Galaxies (SMUDGes; \citealp{Zaritsky_2019}, \citealp{Zaritsky_2022}, \citealp{Zaritsky_2023}) to detect 7070 UDG candidates in the Legacy Survey, spanning over 20,000 square degrees of sky including the Fornax, Virgo, and Coma clusters. 

Catalogs of DGs produced through by-eye image inspection are essential for understanding the reliability and limitations of existing automated detection methods. The first effort to visually construct a relatively complete list of galaxies in the Fornax cluster was the Fornax Cluster Catalog (FCC; \citealp{Ferguson_1989}), identifying 340 likely cluster members of various morphologies including DGs. Visually compiled dwarf catalogs in Fornax have been greatly augmented by two more recent efforts: the Next Generation Fornax Survey (NGFS; \citealp{Munoz_2015}, \citealp{Eigenthaler_2018}, \citealp{Ordenes-Briceno_2018}), which produced a catalog of over 600 dwarf galaxies within the Fornax virial radius, and  \cite{Paudel_2023}, which explored 7643 deg$^2$ of the Legacy Survey to visually catalog early-type dwarf elliptical (dE) galaxies in the local universe (\textit{z} $<$ 0.01), identifying 854 dE galaxies in Fornax. A visual search involving the FDS data has not been conducted since \cite{Venhola_2017}, which yielded 205 dwarfs in the central 4 deg$^2$ of the cluster. Cluster-scale visual catalogs or larger provide a sufficiently representative sample of DGs to pinpoint weaknesses in automated algorithms and can thus constitute a training dataset to improve sensitivity. 

\setcounter{footnote}{0}

Crowdsourcing via citizen science is a growing outlet for rendering manual inspection of large scale surveys a more efficient process, particularly with the Zooniverse web platform\footnote{\url{https://www.zooniverse.org/}}. Zooniverse offers built-in training resources for volunteers to minimize the prior knowledge needed to participate in projects yet still produce quality results. Easily recognizable features of certain galaxies has made morphological classification a popular application of citizen science, most notably by \textit{Galaxy Zoo} \citep{Lintott_2008}. The smooth light profile of DGs in optical images makes them well-suited for identification by those with non-astronomy backgrounds, although complications arise due to factors such as galaxy surface brightness, size, presence of foreground stars, etc. Opening up the research process to volunteers additionally provides an accessible opportunity for the public to learn about extragalactic astronomy while making valuable contributions to scientific projects. 

In this work, we present the first catalog of diffuse galaxies to be identified with Zooniverse in the Fornax cluster, relying on optical imaging data obtained from the Fornax Deep Survey and Legacy Survey. This approach serves as a potentially highly efficient option to map DG populations in large areas of the sky. The observations used in this work and the development of our Zooniverse project are described in Section \ref{sec:search}. Section \ref{sec:galfit} details the process of deriving the photometric parameters of our DG candidates. We present our final catalog of DGs and resulting photometry in Section \ref{sec:results}. The performance of our citizen science approach compared to other automated and visual searches is discussed in Section \ref{sec:discussion}, along with possible insights into the formation history of DGs in Fornax. Throughout the paper we assume the distance to the Fornax cluster is 19 Mpc ($m-M=31.39$ mag) and refer to our final catalog as the \textit{Blurs} catalog.

\section{Citizen Science Search} \label{sec:search}
We conducted our search for DGs in Fornax through a Zooniverse project we developed titled ``Blobs and Blurs: Extreme Galaxies in Clusters". Although our search has concluded, the project URL is still active\footnote{\url{https://www.zooniverse.org/projects/mike-dot-jones-dot-astro/blobs-and-blurs-extreme-galaxies-in-clusters}} and can be accessed to view the project on Zooniverse and the resources we created to teach volunteers about diffuse galaxies. The current interface of the project reflects the most recently completed search conducted in the Virgo cluster \citep{Dey_2025}, which we searched following the completion of the Fornax effort.

\subsection{The Data \label{subsec: data}}
Diffuse galaxies can be detected by-eye based on morphology in a sufficiently deep optical image. The deepest optical survey available spanning the entire Fornax cluster is the Fornax Deep Survey (FDS; \citealp{Peletier_2020}), observed with the OmegaCAM \citep{Kuijken_2002} wide-field camera on the 2.6m VLT Survey Telescope (VST; \citealp{Capaccioli_2012}) in the \textit{u}, \textit{g}, \textit{r}, and \textit{i} bands. We rely on the publicly accessible FDS data (hereafter used synonymously with VST data) to generate optical images needed for classification purposes. 

\begin{figure*}[ht!]
    \centering
    \includegraphics[width=0.75\linewidth]{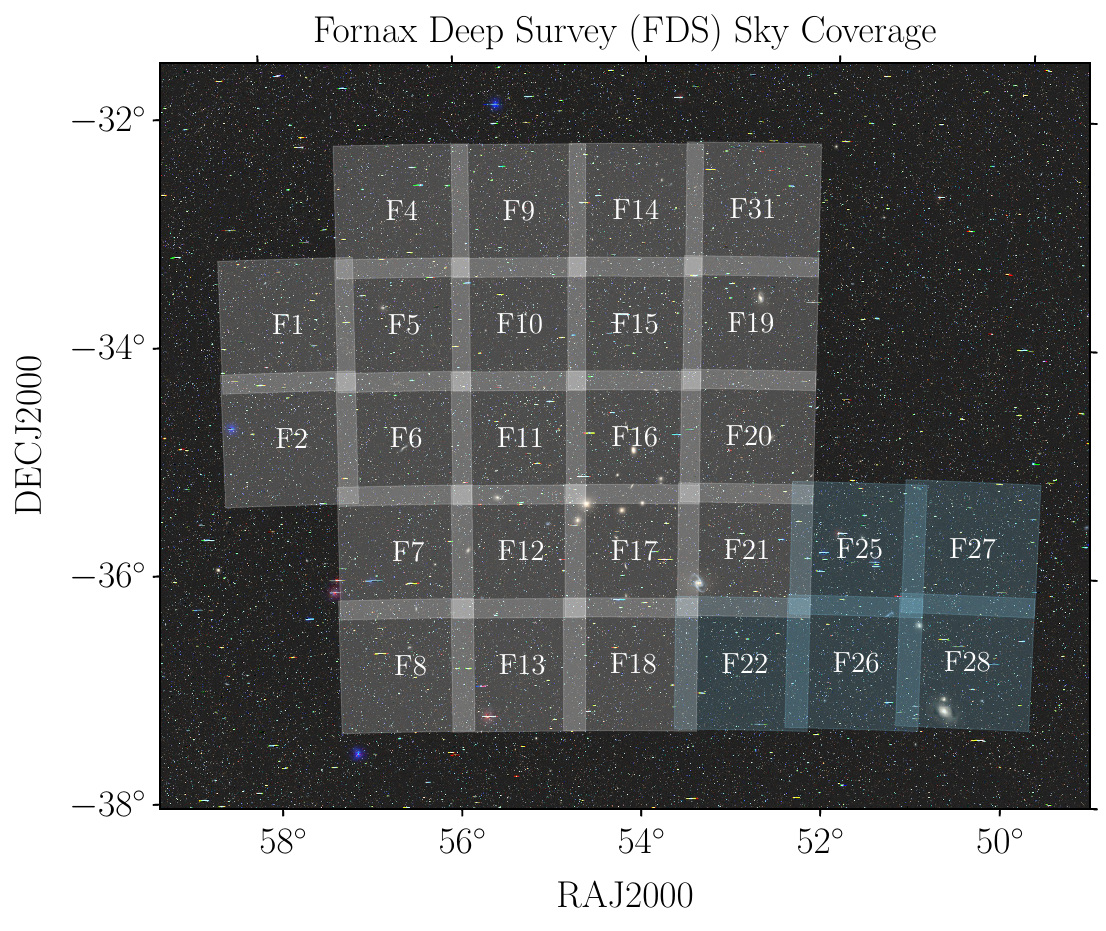}
    \caption{The Fornax Deep Survey (FDS; \citealp{Peletier_2020}) sky coverage plotted over an optical image of the Fornax cluster from the Legacy Survey. Each observing field is numbered and covers an area of approximately 1.17$^{\circ} \times$  1.17$^{\circ}$. The white tinted fields have \textit{ugri} coverage while the blue tinted fields are missing the \textit{u}-band. The F3 and F33 tiles (to the left of F7 and F8) were not used in this project due to incomplete sky and filter coverage.}
    \label{fig:fds_coverage}
\end{figure*}

A customized image stretch for the VST data was carefully selected in order to bring out the faintest DGs. This image stretch was sufficient for identification in most cases except for particularly bright DGs and near bright stars and galaxies that may obscure fainter DGs. We therefore supplemented our VST image set with images from the Dark Energy Camera Legacy Survey (DECaLS hereafter; \citealp{Dey_2019}), sourced from the image cutout service provided by the Legacy Survey Sky Viewer (LSSV).\footnote{\url{https://www.legacysurvey.org/viewer}} Bright stars and the extended halos of bright galaxies in DECaLS are forward-modeled during the sky estimation process to produce visually cleaner images near bright sources. Moreover, the default image stretch of DECaLS cutouts pulled from LSSV does not saturate even for fairly bright DGs. These measures ensure that our inspection process remains robust across all diffuse galaxy demographics and locations in the cluster. 

The FDS divides the Fornax cluster into 28 observing fields numbered 1 through 33\footnote{Fields 23, 24, 29, 30, and 32 of the FDS do not exist and are therefore exceptions to this numbering system.}, each corresponding to an area of $\sim$1 square degree (see Fig. \ref{fig:fds_coverage}). Two of the 28 fields were not used in this work because of insufficient coverage, therefore we examine a total cluster area of 26 deg$^2$. The FDS images tiles were downloaded from the European Southern Observatory (ESO) archive science portal.\footnote{\url{https://archive.eso.org/scienceportal/home}} To generate image cutouts appropriate for the scale of DGs, we adopt a pixel scale of $0.4\arcsec$ per pixel and dimensions of 512$\times$512 pixels. We then split each FDS tile into a grid of 576 cutouts, with slightly overlapping image borders to account for objects cut off by the image frame. We opted to use \textit{ugi} stacked images to produce color cutouts because we were originally also interested in identifying anomalous, isolated star-forming regions (``blue blobs", \citealp{Jones_2022}) that stand out in $u$-band, although ultimately we did not find any good candidates in Fornax. In fields 22, 25, 26, 27, and 28, missing \textit{u}-band data required \textit{gri} stacked images to be used instead. The inclusion of blue blobs in our Zooniverse project required an additional set of image cutouts to be provided to volunteers from the Galaxy Evolution Explorer (GALEX; \citealp{Martin_2005}) (see Section \ref{subsec:blue_blobs}), again sourced from LSSV. The same cutout system used to divide the FDS data was applied to the DECaLS and GALEX data.

\subsection{Zooniverse Workflow\label{subsec: workflow}}
We prepared a brief tutorial for volunteers upon entering the project on Zooniverse (which can still be accessed through the project URL) describing DGs, the set of images they will encounter (ordered by VST, DECaLS, GALEX), and what characteristics to look for when classifying. A wide variety of example classifications and possible image artifacts were provided through a comprehensive field guide, available for consultation. The user interface is shown in Figure \ref{fig:zoon_interface}. To make a classification, volunteers were presented with a subject (i.e., image cutouts from VST, DECaLS, and GALEX covering the same $\sim$3.4\arcmin$\times$3.4\arcmin \ patch of sky) corresponding to a random location in the Fornax cluster and were tasked with completing the following workflow:

\begin{figure*}[ht!]
    \centering
    \includegraphics[width=0.75\linewidth]{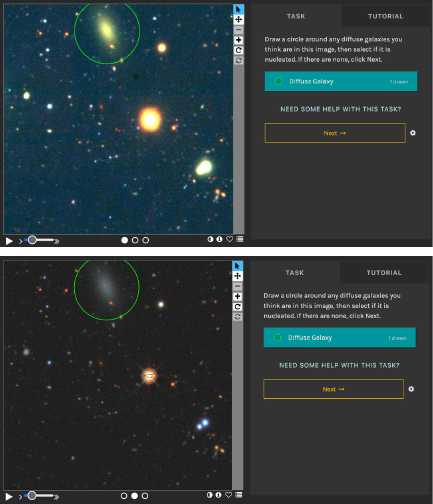}
    \caption{Zooniverse interface used by volunteers to classify objects on our project. The top screenshot shows the user interface when the VST image is selected and the bottom screenshot is the same interface with the DECaLS image selected. The three circles below the image cutout allow the volunteer to select from the three image types we provide: VST (left circle), DECaLS (middle circle), and GALEX (right circle). In this cutout, a nucleated diffuse galaxy is present in the field of view, therefore a volunteer would be expected to draw a circle around it with the Zooniverse shape tool. After drawing the circle, a pop-up window asking if this object is nucleated will appear, which in this case they should select ``Yes".}
    \label{fig:zoon_interface}
\end{figure*}

\begin{itemize}
  \item \textbf{Task 1:} Draw a circle around any diffuse galaxies you think are in this image, then select if it is nucleated. If there are none, click Next.
  \item \textbf{Subtask of Task 1:} Is this diffuse galaxy nucleated? (yes/no)
  \item \textbf{Task 2:} Draw a box around any blue blobs you think are in this image. If there are none, click Done.
\end{itemize}

\vspace{-0.2cm}
To complete the project in a reasonable amount of time, cutouts were retired from the subject pool once 10 classifications were recorded. Instances where a volunteer dismissed a subject, without marking any objects, counted as a classification. In fact, these likely constitute the majority of classifications as many cutouts contain no DGs. Participants were given the option to post any subject to an internal discussion board to consult the project scientists about the image content. We suggested volunteers classify DGs that appear no smaller than 0.5 cm in diameter, which on a standard desktop screen corresponds to an angular diameter of roughly 10$\arcsec$, with the intention of avoiding excessive contamination from background galaxies. In practice, this suggestion is difficult to enforce because the visible extent of a DG is often ill-defined and most users will not aim to be this precise. We recognize that this complicates our completeness for small galaxies and defer to Sections \ref{subsec:comp} and \ref{subsec:final_cut} to discuss this in detail. 

\begin{figure*}[ht!]
    \centering
    \includegraphics[width=0.6\linewidth]{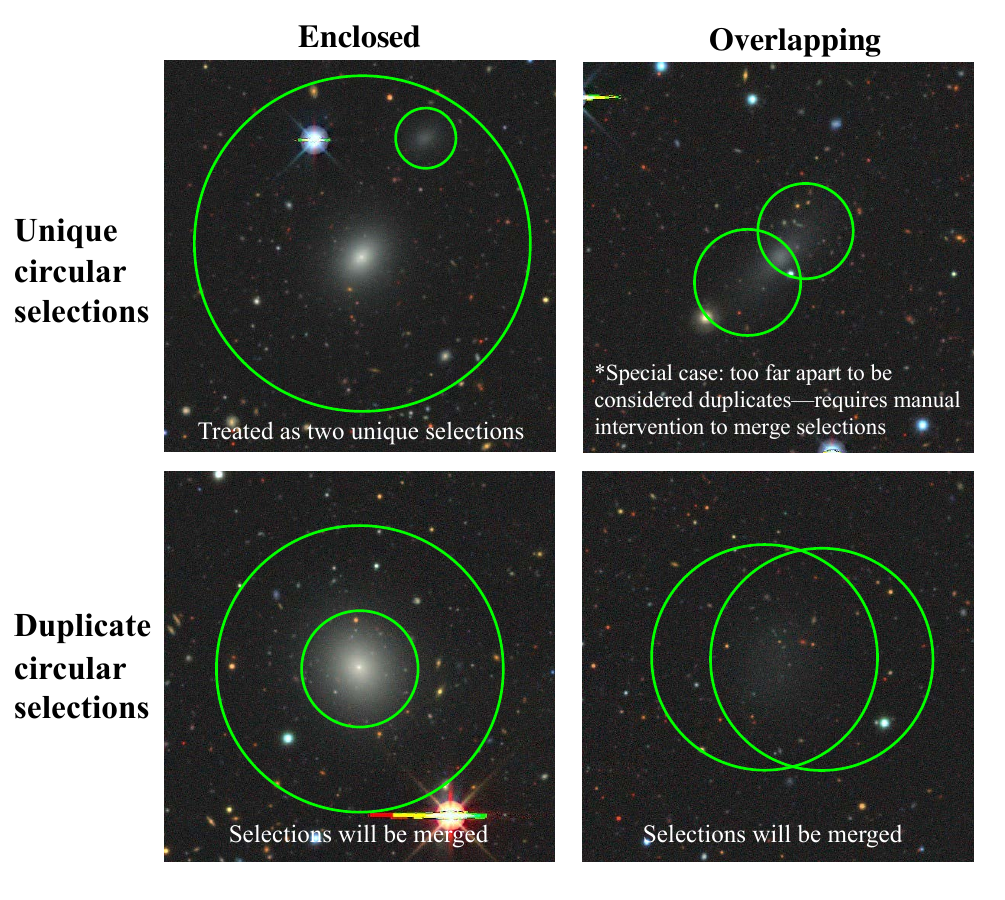}
    \caption{Visual aid demonstrating how unique and duplicate diffuse galaxy selections are distinguished by our reduction process. The green circles indicate hypothetical selections of DGs that could realistically be drawn by Zooniverse volunteers on DECaLS cutouts when a DG is present (intentionally centered on object). Two (or more) circular selections are considered unique objects if their respective centers fall outside of the area in which the circles intersect. A special case occurs (top right panel) when volunteers do not center their selection on a galaxy, possibly because the object is quite large or it is split across multiple cutouts. In this case, the selections must be manually merged together.}
    \label{fig:dupl_guide}
\end{figure*}

 \subsection{Catalog Preparation \label{subsec: reduc}}
 \textit{Blobs and Blurs} was launched to Zooniverse in June 2023 with a total of 14,976 image cutouts to examine, spanning 26 deg$^2$ of Fornax. All of these cutouts were retired in approximately 30 days due to participation from over 1,400 volunteers, resulting in 149,760 classifications. Circular selections (for DGs) were drawn a total of 27,979 times across all cutouts, of which 15,153 ended up being unique. We merged duplicate selections of the same object if the intersection of both circles contained their respective centers (see Fig. \ref{fig:dupl_guide}). Within the same cutout, only selections made by different users qualified as a duplicate detection to counteract excessive drawings made by the same user (possibly artificial). Relying on classifications made by citizen scientists naturally introduces some ambiguity into the data, as volunteers might differ significantly in their judgment of an object’s center and the extent of its radius when circling a potential DG. This mostly impacts larger DGs or those split across multiple cutouts, resulting in some instances where duplicate detections were manually merged together, though this proved to only represent a small fraction of cases. Thus, applying a maximum separation tolerance to merge duplicates alone is ineffective for handling the range of scatter in the selection centroid. We set a minimum detection threshold, $N_{\mathrm{min}}$, of 4 for an object to be considered a diffuse galaxy candidate and cut all objects with fewer detections than this, leaving 1,203 candidates out of 15,153. This step subsequently removed any accidental or artificial DG classifications from the detection list, as well as the vast majority of unlikely candidates.

\subsection{Quality Flags \label{subsec:comp}}
Incorrectly identified objects may still exist in the candidate pool even after setting a minimum detection threshold because of commonly made mistakes by volunteers when classifying, such as confusing a diffuse galaxy with another morphology, image artifacts, or background galaxies. To further filter out these objects, we cross-matched our list of DG candidates with existing dwarf catalogs of Fornax. The FDS-produced catalogs by \cite{Venhola_2018} and \cite{Venhola_2022} (hereafter V18 and V22) offer the most comprehensive comparison for our candidates list providing a combined total of $\sim$800 dwarf galaxies found through automated detection. \cite{Venhola_2018} additionally provides a catalog of over 13,000 background galaxies identified in Fornax as a result of cuts used to assess cluster membership. We additionally compare to the Fornax portion of the SMUDGes catalog (\citealp{Zaritsky_2019}, \citealp{Zaritsky_2022}) and dwarfs found by the Next Generation Fornax Survey (NGFS; \citealp{Munoz_2015}, \citealp{Eigenthaler_2018}, \citealp{Ordenes-Briceno_2018}), \cite{Paudel_2023} (P23 hereafter), and the Fornax Cluster Catalog (FCC; \citealp{Ferguson_1989}), which provide $\sim$200, 650, 400, and 200 objects to cross-match against, respectively. 

Candidates from our list of 1,203 which matched with any object in the Fornax dwarf catalogs (V18, V22, SMUDGes, FCC, NGFS, and P23) immediately moved onto the photometry stage of our final catalog preparation. We reviewed all of the remaining objects from our sample by-eye and compared with the V18 background galaxies catalog to eliminate objects most likely to be false positives. DG candidates not found in Fornax by these six dwarf catalogs (or any existing catalog) were identified during this review process as well. These series of cuts reduced our preliminary list from 1,203 to 714 candidates. We manually inspect each of these 714 objects ourselves while performing the catalog photometry discussed in Section \ref{subsec:final_cut}, which is the last cut made to produce our final catalog.   

\subsection{Search for Blue Blobs \label{subsec:blue_blobs}}
We note that our Zooniverse project did not focus on Fornax diffuse galaxies alone. We simultaneously searched for isolated blue stellar systems known as blue blobs within the same image set, which have only been discovered in the Virgo cluster thus far (see \citealp{Jones_2022}). Blue blobs are thought to be the ram pressure stripping equivalents of tidal dwarf galaxies but are too isolated to confidently identify any parent galaxy. These objects exhibit active star formation with strong emission in both the UV and H$\alpha$, motivating the addition of GALEX images to our project. No blue blobs were detected in Fornax from our search, therefore we reserve our primary discussion in this paper for DGs. Our search in the Virgo cluster, however, discovered 34 new blue blob candidates (see \citealp{Dey_2025}). This Virgo data set will be similarly analyzed in future work to study the DGs. 

\section{Deriving Physical Parameters with GALFIT}{\label{sec:galfit}}
\subsection{Fitting Pipeline}
To perform statistical analyses of the DGs found in Fornax, we obtain photometry measurements of the remaining 714 DG candidates with the 2D model-fitting package GALFIT \citep{Peng_2010}. We adopt a similar fitting pipeline to that established by \cite{Khim_2024}, which fit objects for the SMUDGes catalog using images from the Legacy Survey. This approach employs a two-stage multicomponent fitting procedure capable of fitting extended sources with one or more S\'ersic profiles and a point-spread function (PSF) profile, if an NSC exists. The best-fit model is selected according to the Akaike information criterion (AIC; \citealp{Akaike_1974}), which allows for models with different degrees of freedom to be compared fairly. The model output parameters consist of the central object position, effective radius ($r_{\mathrm{eff}}$), apparent magnitude ($m$), S\'ersic index ($n$), axis ratio ($b/a$), position angle (PA), central and effective surface brightness ($\mu_0$ and $\mu_e$), and flat sky background level. For our catalog, we derive these parameters from the FDS imaging in the \textit{g} and \textit{r} bands, providing GALFIT with a 500$\times$500 pixel image centered on each target (0.2$\arcsec$ pixel scale), a constructed PSF model, flux uncertainty image ($\sigma$-image), a bright pixel mask, and a constraint file specifying the free parameter space. We refer to \cite{Khim_2024} for a more detailed description of the automatic masking schemes and model components used at each fitting stage. 

We uniformly fit all of the candidates in the same initial batch then iteratively grouped and re-fit the poorly fit targets in progressively smaller batches. These batches were organized based on shared factors to which GALFIT is sensitive to, such as the presence of an NSC or bright contaminants, allowing the masking strength to be adjusted accordingly. For the smallest batches, the degree of masking was tailored to each target individually. In cases where the automatic masking could not be further increased without covering significant portions of the target, excessively bright pixels were manually masked via regions created with \texttt{SAOImage DS9} \citep{SAODS9}. If the extended halos of bright contaminant objects in the FDS data could not be reasonably masked with any method, we opted to fit these candidates with equivalent images and PSFs from DECaLS, as these bright sources are more effectively mitigated by the DECaLS processing. Ideally, all candidates would be fit with the FDS data because it is a deeper optical survey than DECaLS. For all 714 candidates, we visually inspected the GALFIT model (as selected by the AIC value) and its residuals before declaring a fit acceptable. In some cases, we declared the simple S\'ersic model the best-fit model because it produced a better looking fit than the AIC-selected model, which was typically the case for objects with more difficult fitting circumstances. We find that our photometry is in good agreement with the photometric parameters reported by V18 and SMUDGes, as they also perform their catalog photometry with GALFIT in both the \textit{g} and \textit{r} bands (see Appendix \ref{appendix:galfit}).

\subsection{PSF Creation}
We generated \textit{g} and \textit{r} band PSFs for each respective tile of the FDS by first locating point sources with the DAOFIND algorithm \citep{Stetson_1987} contained in the \texttt{photutils} package. The best quality point sources were manually selected to construct each PSF. We compared the output full width at half maximum (FWHM) to the FWHM listed in the tile header from the FDS observations, ranging from 0.83-1.39$\arcsec$ in \textit{g} and 0.69-1.42$\arcsec$ in \textit{r}. In general, we find that our FWHM values are consistent with the values from the FDS, differing at most by 1-2 pixels (\textit{g} range: 0.79-1.28$\arcsec$; \textit{r} range: 0.76-1.16$\arcsec$). 

\subsection{Flux Uncertainty Image Creation}
The pixel-to-pixel uncertainties in flux, including the Poisson noise, are quantified by GALFIT through the $\sigma$-image. We constructed $\sigma$-images using the weight maps produced by \cite{Venhola_2018} according to the equation: 

\begin{equation}\label{eq:sigma}
    \sigma = \left( \frac{0.2}{0.21} \right)^2 \sqrt{\frac{1}{W} + \frac{f}{\text{GAIN} \sqrt{N}}}
\end{equation}

\noindent where $W$ is the weight map pixel value and $f$ is the unmodified flux value from the science image in ADUs per second. GAIN and $N$ correspond to the effective gain and number of stacked raw data files, read in directly from the FDS tile header. The factor of $(0.2/0.21)^2$ accounts for the transition in the FDS processing from an instrumental pixel scale of $0.21\arcsec$ per pixel to $0.2\arcsec$ per pixel in the final science image (\citealp{Kuijken_2002}; \citealp{Su_2021}).  

\subsection{Final Cut of Unlikely Candidates \label{subsec:final_cut}}
While inspecting the GALFIT residuals of each target, we visually classified each candidate ourselves to more expertly determine if an object is nucleated and flag any lingering objects unlikely to be a DG (e.g., other morphology or background source). We use the values for effective radii ($r_{\mathrm{eff}}$) measured by GALFIT as a proxy for the $\sim$10$\arcsec$ minimum diameter we suggested volunteers follow when circling DGs (diameter of drawn circle on image). We therefore treat $r_{\mathrm{eff}}=$ 5$\arcsec$ roughly as the lower size limit for our search, below which our completeness becomes less reliable. After inspecting all 714 objects, the flagged candidates with $2r_{\mathrm{eff}} < 10\arcsec$ were considered likely background galaxies and cut from our sample, whereas those with $2r_{\mathrm{eff}} \ge 10\arcsec$ were further evaluated by N. Mazziotti, M. Jones, and D. Sand and also assessed for cluster membership based on recession velocities available on \texttt{SIMBAD} \citep{SIMBAD}. Objects receding at a rate of more than $\sim$3000 km s$^{-1}$, approximately double the mean recession velocity of galaxies in the Fornax cluster \citep{Drinkwater_2001}, were deemed background to Fornax and dropped. Flagged objects larger than $10\arcsec$ without a recession velocity available were cut only if they were judged to be morphologically distinct from a diffuse galaxy.
 
\begin{figure*}[ht!]
    \centering
    \includegraphics[width=0.75\linewidth]{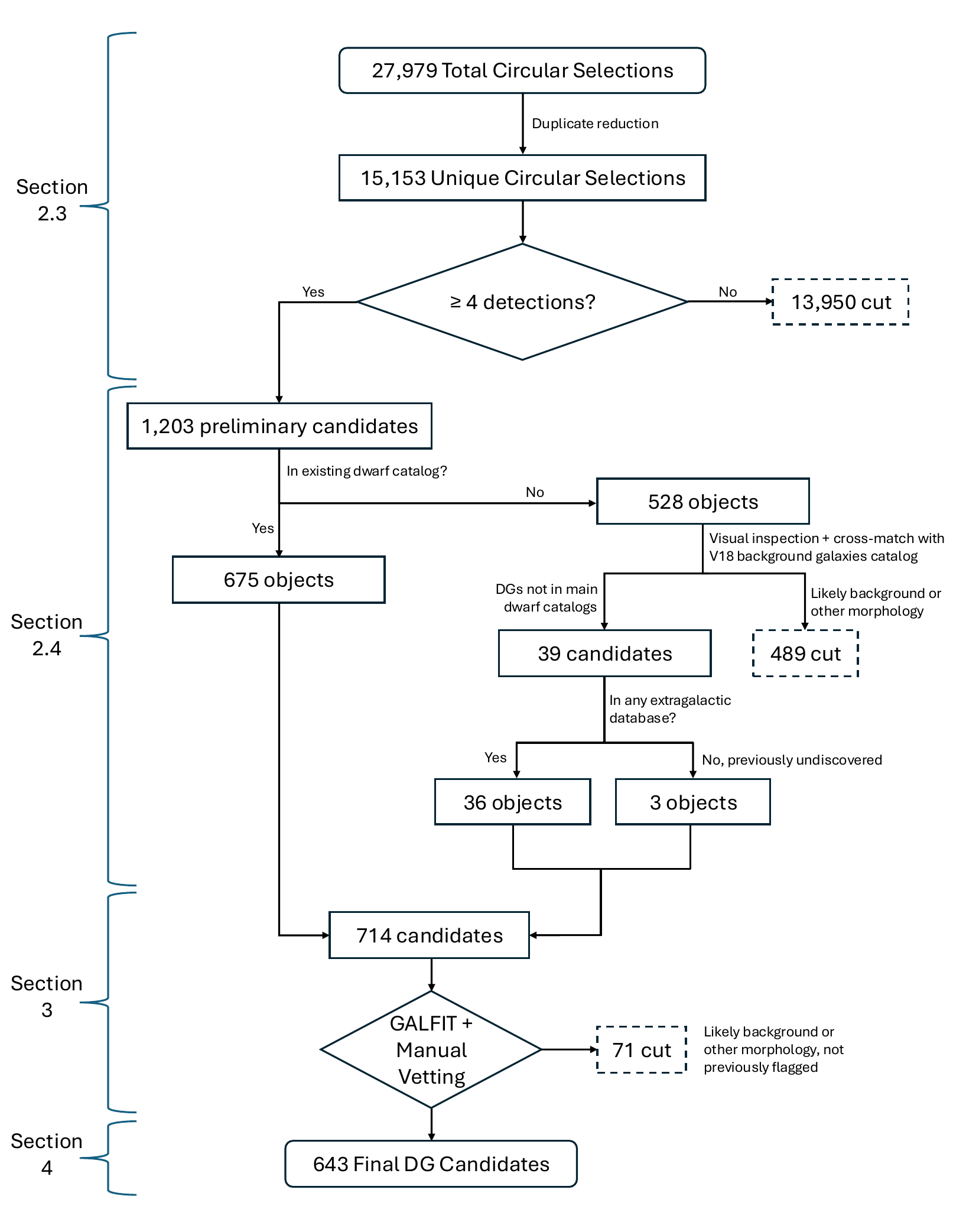}
    \caption{Flowchart demonstrating the successive cuts made to reduce the initial list of 27,979 circular selections drawn by volunteers to a final catalog of 643 DG candidates. Information on each stage of the catalog preparation process can be found in the corresponding sections indicated by the curly brackets. 
    }
    \label{fig:flow_chart}
\end{figure*}

\section{Final Catalog}{\label{sec:results}
The cuts described above are summarized in flowchart format and shown in Figure \ref{fig:flow_chart}, leaving us with a final catalog of 643 diffuse galaxy candidates, which we hereafter refer to as the Blurs catalog. A complete table of these objects, including our GALFIT estimates of their photometric parameters, is available in machine-readable format (see Table \ref{tab:column_descr}). To assess the completeness of Blurs, we examine the parameter space in which we recover faint galaxies from previous automated and visual searches of Fornax in Sections \ref{subsec:auto_comp} and \ref{subsec:vis_comp}, respectively. The statistics describing our completeness relative to these searches are listed in Table \ref{tab:comparison_summary}. All comparisons discussed below are limited to the region in Fornax where our catalog and the comparison sample fully overlap. Objects in Blurs that were not identified by any of the automated catalogs are highlighted in Appendix~\ref{appendix:weFound_NotAutomated} and discussed in Sections \ref{subsec:auto_comp} and \ref{discussion:automated_missed}. We display DECaLS images of the objects we do not recover from these catalogs in Appendix~\ref{appendix:not_recovered} to evaluate whether their morphologies align with our search criteria. We refer to Section \ref{discussion:recovery_success} for a broader discussion on how these metrics constrain the overall sensitivity of our catalog.

\begin{deluxetable*}{lll}
\tabletypesize{\footnotesize}
\tablewidth{0pt}

\tablecaption{ Description of Blurs Final Catalog \label{tab:column_descr}}

\tablehead{
\colhead{Column Name} & \colhead{Description} & \colhead{Format/Units} 
}

\startdata 
Target & Name of object & BLUR designator followed by       coordinates \\
RA & Right ascension (J2000) & Degrees \\
Dec & Declination (J2000) & Degrees \\
$r_\mathrm{eff}$ & Effective radius & Arcseconds \\
$\sigma_{r_\mathrm{eff}}$ & 1$\sigma$ GALFIT uncertainty in effective radius & Arcseconds \\
\textit{n} & Sérsic index & Unitless \\
$\sigma_n$ & 1$\sigma$ GALFIT uncertainty in Sérsic index & Unitless \\
$b/a$ & Axis ratio & Unitless \\
$\sigma_{b/a}$ & 1$\sigma$ GALFIT uncertainty in axis ratio & Unitless \\
PA & Position angle of major axis & Degrees \\
$\sigma_{\mathrm{PA}}$ & 1$\sigma$ GALFIT uncertainty in position angle & Degrees \\
$m_g$ & Apparent magnitude in \textit{g}-band  & AB Mag \\
$\sigma_{m_g}$ & 1$\sigma$ GALFIT uncertainty in \textit{g}-band apparent magnitude & AB Mag \\
$\mu_e^g$ & Surface brightness in \textit{g}-band at effective radius  & mag/arcsec$^2$ \\
$\mu_0^g$ & Central surface brightness in \textit{g}-band & mag/arcsec$^2$ \\
$\sigma_{\mu_0^g}$ & Central surface brightness in \textit{g}-band & mag/arcsec$^2$ \\
$m_r$ & Apparent magnitude in \textit{r}-band  & AB Mag \\
$\sigma_{m_r}$ & 1$\sigma$ GALFIT uncertainty in \textit{r}-band apparent magnitude & AB Mag \\
$\mu_e^r$ & Surface brightness in \textit{r}-band at effective radius  & mag/arcsec$^2$ \\
$\mu_0^r$ & Central surface brightness in \textit{r}-band & mag/arcsec$^2$ \\
$\sigma_{\mu_0^r}$ & Central surface brightness in \textit{r}-band & mag/arcsec$^2$ \\
Nucleated & Nucleation of object & 0 = non-nucleated, 1 = nucleated, -1 = uncertain \\
$m_{g,\mathrm{NSC}}$ & Apparent magnitude of NSC in \textit{g}-band & AB Mag \\
$\sigma_{m_{g,\mathrm{NSC}}}$ & 1$\sigma$ GALFIT uncertainty in apparent magnitude of NSC in \textit{g}-band & AB Mag \\
$m_{r,\mathrm{NSC}}$ & Apparent magnitude of NSC in \textit{r}-band & AB Mag \\
$\sigma_{m_{r,\mathrm{NSC}}}$ & 1$\sigma$ GALFIT uncertainty in apparent magnitude of NSC in \textit{r}-band & AB Mag \\
$\Delta_{NSC}$ & Angular separation of NSC from Sérsic model center & Arcseconds \\
$\log{M_\ast}$ & Stellar mass of galaxy & log($M_*$/$M_{\odot}$) \\
$\log{M_{\mathrm{NSC}}}$ & Stellar mass of NSC & log($M_{\mathrm{NSC}}$/$M_{\odot}$) \\
Reference & Catalogs where this object appears & Referenced catalogs: [V18, V22, SMUDGes, NGFS, P23, FCC, Other] \\
\enddata
\tablecomments{The 1$\sigma$ uncertainties for $\mu_e^g$ and $\mu_e^r$ are equivalent to the corresponding uncertainties for $m_g$ and $m_r$. Candidates that only appear in an extragalactic database and not in any of the six named dwarf catalogs we compare to are labeled as ``Other" under the ``Reference" column. }
\end{deluxetable*}

\begin{deluxetable*}{lccccc}
\tablecaption{Catalog Comparison Summary \label{tab:comparison_summary}}
\tablewidth{0pt}
\tablehead{
\colhead{Catalog (\# of objects)} & 
\colhead{\% of catalog we recover} & 
\colhead{\% recovered with $r_{\mathrm{eff}} \ge 5\arcsec$} &
\colhead{Objects not recovered from \textit{Blurs}}
}
\startdata
\multicolumn{4}{c}{\textbf{Automated Catalog Comparison Summary}} \\
\hline
V18 (541)       & 76.2\% (412/541) & 88.4\% (312/353) & 231 \\
V22 (261)       & 42.9\% (112/261) & 65.2\% (43/66) & 531 \\
SMUDGes (198)   & 98.0\% (194/198) & 98.0\% (194/198) & 449 \\
\hline
\multicolumn{4}{c}{\textbf{Visual Catalog Comparison Summary}} \\
\hline
NGFS (659)      & 72.2\% (476/659) & 80.6\% (345/428) & 167 \\
P23 (416)       & 95.4\% (397/416)  & 96.3\% (262/272) & 246 \\
FCC[dE] (197)   & 97.0\% (191/197)  & 98.3\% (178/181) & 452 \\
\enddata
\tablecomments{The completeness of Blurs compared against existing automated and visual catalogs of dwarf/faint galaxies in Fornax, without considering possible differences in the photometric parameter space explored by each search. Some objects in each catalog were recovered by us but ultimately did not make our final catalog cut and are therefore considered not recovered. When limiting each catalog to objects with $r_{\mathrm{eff}} \ge 5\arcsec$, the recovery fraction exceeds 80\% for all catalogs except V22. We only compare with objects from each catalog that exist within the sky coverage we examined.}
\end{deluxetable*}

\subsection{Previously Undetected Diffuse Galaxies}
Within the small/faint area of our catalog parameter space, we identify three diffuse galaxy candidates that have not been detected by any existing searches of Fornax, automated or visual, nor do they appear in any extragalactic database. These candidates are presented in Figure \ref{fig:new_cands}. Although all three objects could conceivably be background to Fornax, no recession velocities are currently available in the literature to rule out these galaxies as such. Regardless of whether these candidates belong to the cluster, they may have gone undetected because all three are located either in the vicinity of bright stars, by the border of the FDS coverage, or both. Previously conducted searches of Fornax might not have inspected (or possibly masked) the areas where these objects are found, or were more insensitive to this regime of small/faint objects than our visual search. 

\begin{figure*}[ht!]
    \centering
    \includegraphics[width=0.75\linewidth]{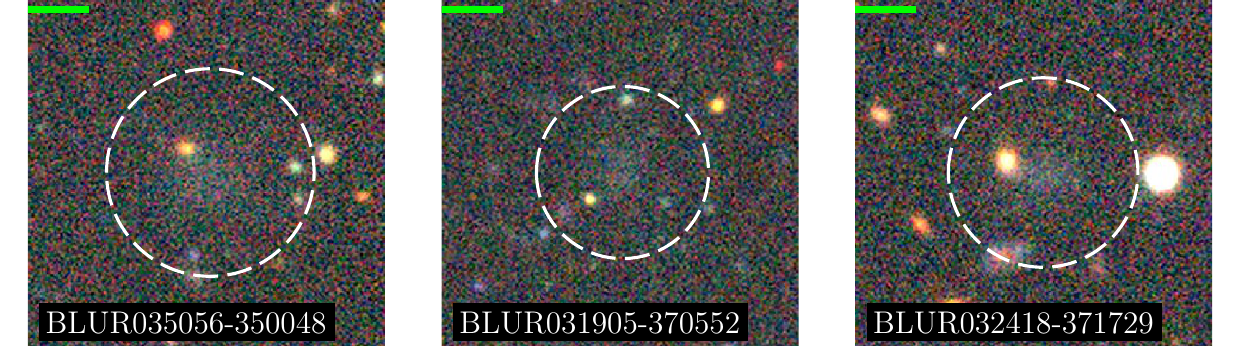}
    \caption{Three potential diffuse galaxies (indicated by dashed circle) detected by Blurs that are not present in any existing catalogs of Fornax, automated or visual. The green bar at the top of each image represents 10\arcsec in the image pixel scale. All images are DECaLS cutouts taken from the Legacy Survey Sky Viewer.}
    \label{fig:new_cands}
\end{figure*}

\subsection{Comparison to Automated Catalogs \label{subsec:auto_comp}}
In this subsection, we focus on objects from V18, V22, and SMUDGes that are recovered by Blurs, while objects identified by Blurs but missed by these automated searches are primarily discussed in Section \ref{discussion:automated_missed}. The V18, V22, and SMUDGes catalogs were produced through a variety of automated detection strategies optimized for finding faint galaxies. Both V18 and SMUDGes implement Source Extractor to prepare their initial detection list, which has become a widely used tool for targeting groups of pixels above a desired brightness threshold. SMUDGes additionally employs a convolutional neural network, EfficientNetB1 \citep{Tan_2019}, to assign classifications to their detected objects, achieving an overall accuracy of 96.2\% \citep{Zaritsky_2022}. In the cluster environment, where the number of objects is especially dense, the sensitivity of algorithms like Source Extractor to LSB galaxies can be significantly reduced. The goal of the V22 catalog was to increase the completeness of V18 in the LSB regime, relying on a max-tree based object detection algorithm that assembles a hierarchy of sources according to brightness and overlapping pixels, preventing fainter objects from being rejected due to a fixed, global threshold (as is the case for Source Extractor). Since we rely on the same FDS data used by both V18 and V22, as well as the Legacy Survey imaging used by SMUDGes, understanding where our methodology is less sensitive, equivalent, or perhaps preferred to these algorithms is important for identifying how automated detection of DGs can be improved. 

We display the GALFIT estimates of the Blurs objects in size-luminosity space (effective radius $r_{\mathrm{eff}}$ versus absolute \textit{r}-band magnitude $M_r$) in Figure \ref{fig:rMag_reff}. To determine where in this parameter space our catalog is more insensitive than automated detection, we additionally include in this plot the objects we do not recover from V18, V22, or SMUDGes. To highlight where these catalogs are less sensitive than Blurs, the objects they do and do not recover from us are plotted in size-luminosity space in Figure \ref{fig:ext_coverage}. 

\begin{figure*}[ht!]
    \centering
    \includegraphics[width=0.5\linewidth]{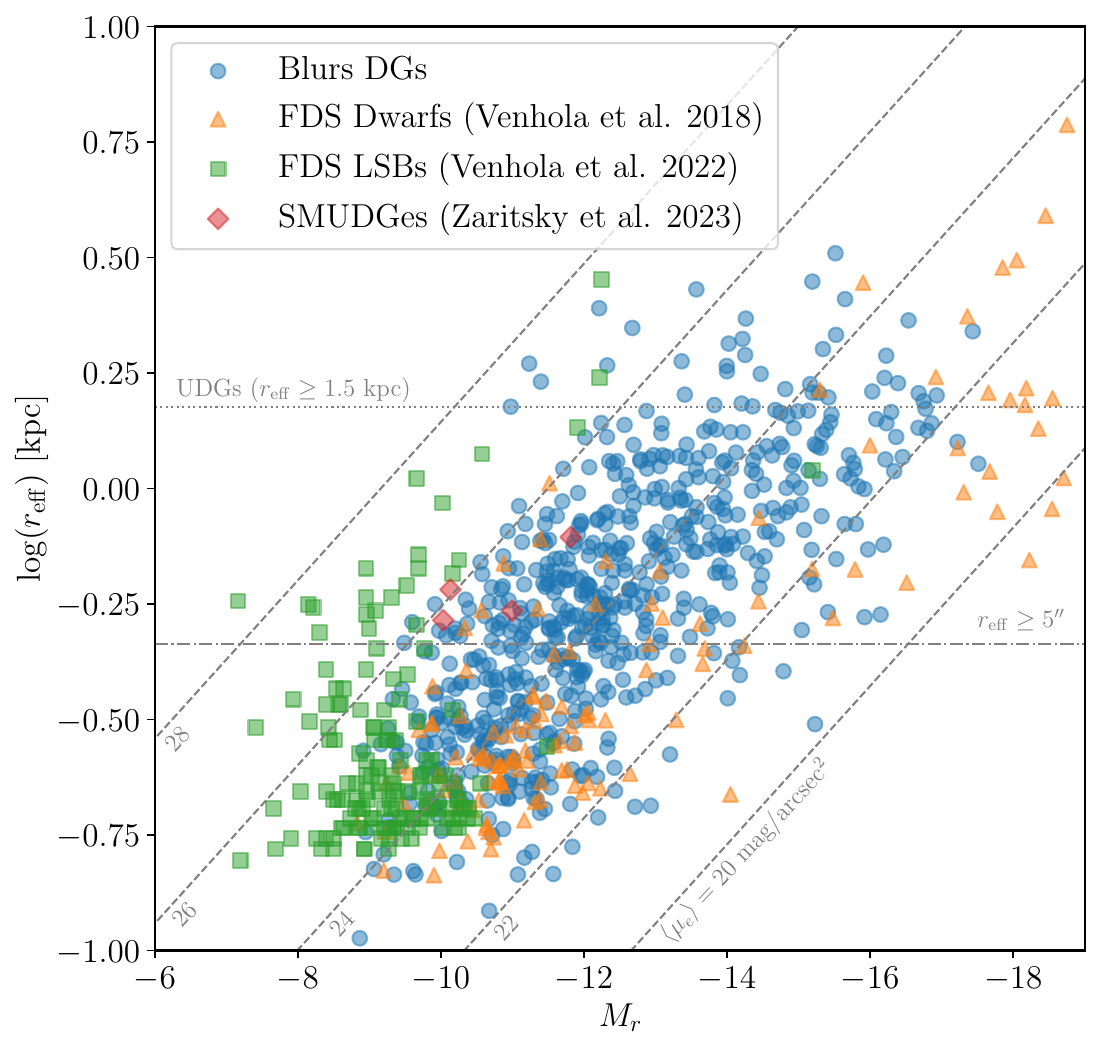}
    \caption{Properties of Blurs DGs in size-luminosity space, plotted for effective radius $r_{\mathrm{eff}}$ (log of $r_{\mathrm{eff}}$ in kpc, assuming a distance to the Fornax cluster of D=19 Mpc) and absolute magnitude $M_r$, both in the \textit{r} band. The objects we did not recover from the three automated catalogs discussed in this work—V18, V22, and SMUDGes—are indicated by the orange triangles, green squares, and red diamonds, respectively. The UDG threshold of $r_{\mathrm{eff}} \ge 1.5$ kpc is denoted by the horizontal dotted line, while the dash-dotted line represents the minimum size limit of $r_{\mathrm{eff}} = 5\arcsec$ we suggested volunteers follow. Diagonal dashed lines correspond to constant effective surface brightness, becoming fainter as one goes from the bottom right corner to the top left. We achieve a high level of completeness throughout the entire parameter space except for small, faint objects.}
    \label{fig:rMag_reff}
\end{figure*}

\begin{figure*}[ht!]
    \centering
    \includegraphics[width=1.0\linewidth]{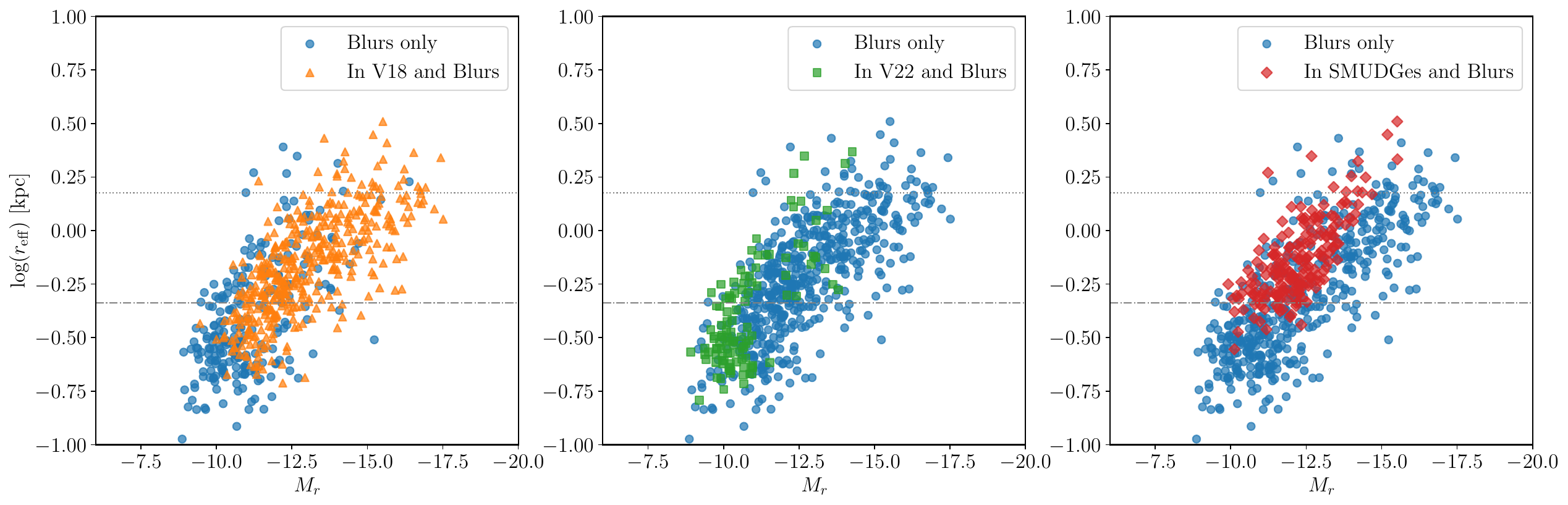}
    \caption{Objects in Blurs not recovered by V18 (left panel), V22 (middle panel), or SMUDGes (right panel) in size-luminosity space. Within the parameter space shared by V18 and SMUDGes with Blurs, we find that our catalog is more complete. V18 was intentionally limited to objects with $r_{\mathrm{eff}} > 0.2$ kpc, whereas SMUDGes was restricted to objects with $r_{\mathrm{eff}} \ge 5.3\arcsec$ and a central surface brightness in the \textit{g} band of $\mu_{0}^{g} > 24$ mag arcsec$^{-2}$. The horizontal dotted and dash-dotted lines correspond to an $r_{\mathrm{eff}}$ of 1.5 kpc (UDG threshold) and 5\arcsec as in Fig. \ref{fig:rMag_reff}.}
    \label{fig:ext_coverage}
\end{figure*}

\subsubsection{V18 and V22 Comparison}
The V18 catalog contains dwarf galaxies classified as one of the following three morphological types: early-type dwarfs with a smooth red appearance, early-types with internal structures such as a bulge, bar, or spiral arms, and blue late-type dwarfs with star-forming clumps. The smooth early-type dwarfs most closely fit the description of DGs we asked Zooniverse volunteers to identify. We recover 412 of the 541 total dwarfs from V18 (76.2\%), leaving 129 objects in V18 undetected. In Fig. \ref{fig:rMag_reff}, these 129 objects (orange triangles) are mostly below our effective size limit of $r_{\mathrm{eff}} = 5\arcsec$. Only 19 of the 129 unrecovered objects have an $r_{\mathrm{eff}} \ge 5\arcsec$ and are classified as smooth early-type dwarfs, therefore the remaining 110 missed objects do not align with the type of galaxies targeted by our search; for example, the large objects that were missed are mostly blue late-type dwarfs. When considering only the smooth early-types of V18 with $r_{\mathrm{eff}} \ge 5\arcsec$, we achieve an excellent recovery fraction of 93.9\% (294/313). We are less complete for $r_{\mathrm{eff}} < 5\arcsec$ as expected, recovering only 35.0\% (48/137) of the smooth early-types. Conversely, Blurs contains 231 DGs not identified in V18. In Fig. \ref{fig:rMag_reff}, we observe a diagonal band of candidates in Blurs near or fainter than $\langle \mu_r \rangle \sim$25 mag/arcsec$^2$ that V18 did not recover, suggesting that our catalog overall reaches a fainter surface brightness limit (though V22 addresses this). We also identify a surprising amount of candidates smaller than the $r_{\mathrm{eff}}$ cutoff used in V18 of 0.2 kpc ($2.17\arcsec$), despite focusing our search on objects with $r_{\mathrm{eff}} \ge 5\arcsec$. 

V22 is not classified by morphology in the same way as V18 so we opt to compare against the entire catalog of 261 objects. We identified 112 of these LSB galaxies from V22 (42.9\%), 61.6\% of which have $r_{\mathrm{eff}} < 5\arcsec$, and failed to detect 149, 82.5\% of which have $r_{\mathrm{eff}} < 5\arcsec$ and $M_r >$ -10.4. This gap in completeness in effective radius is unsurprising given that the majority of these objects are much smaller than what we expected volunteers to classify. In Fig. \ref{fig:rMag_reff}, there is a clear surface brightness limit above 26 mag arcsec$^{-2}$ where the objects we miss from V22 exist (green squares). Since V22 was intentionally limited to LSB galaxies to compliment V18, many of the DGs in Blurs were not recovered by V22 (middle panel of Fig. \ref{fig:ext_coverage}). 

\subsubsection{SMUDGes Comparison}
We find excellent agreement with the SMUDGes catalog, detecting 194 (98.0\%) of the objects in SMUDGes without imposing any restrictions on $r_{\mathrm{eff}}$. There are only four objects in SMUDGes missed by Blurs that are represented by the four red diamonds in Figure \ref{fig:rMag_reff}: all four objects have an effective surface brightness near $\sim$26 mag arcsec$^{-2}$, which is the same limit we observe with the V22 objects that we did not recover. One of the four objects more closely resembles a late-type dwarf morphology not targeted by our search, therefore our relative completeness is only reduced by the remaining three objects. Over 400 objects in Blurs were not detected by SMUDGes, as shown by the right panel of Fig. \ref{fig:ext_coverage}, although this is largely because our search extends well past the parameter space of SMUDGes. This is a result of \cite{Zaritsky_2022} defining UDGs as objects with $\mu_{0}^{g} \ge$ 24 mag arcsec$^{-2}$ and $r_{\mathrm{eff}} \ge$ 5.3\arcsec, corresponding to the angular size of the first UDG surveys conducted in the Coma cluster (\citealp{Koda_2015}; \citealp{Dokkum_2015}; \citealp{Yagi_2016}), in order to detect similarly sized objects across the sky. Within the same parameter space as SMUDGes, we detect 67 objects that SMUDGes did not identify, 9 of which are UDG-class  ($r_\mathrm{eff} \ge$ 1.5 kpc). 

\subsection{Comparison to Visual Catalogs \label{subsec:vis_comp}}
In this subsection, we perform a comparison with the Fornax dwarf catalogs produced through visual inspection: the NGFS, \cite{Paudel_2023} (P23), and FCC. We are unable to photometrically assess our relative completeness to these visual catalogs in the \textit{r}-band because they use different bands for their photometry. Nonetheless, we are still able to perform a cross-match comparison and use a combination of the $g$ and $i$ band parameters these catalogs do provide to roughly infer our completeness in size-luminosity space in the \textit{r}-band. We focus our comparison on NGFS and P23, but we note that Blurs identifies 191 (97.0\%) of the dE galaxies in the FCC, missing only 6, and significantly expands upon the FCC completeness by identifying 452 DGs unique to Blurs. 

The three separate NGFS releases were combined into one catalog of 659 dwarf galaxies that covers most of the Fornax cluster (\citealp{Munoz_2015}; \citealp{Eigenthaler_2018}; \citealp{Ordenes-Briceno_2018}). We recover 476 (72.2\%) and miss 183 objects across the full catalog, but recover 80.6\% of objects with $r_{\mathrm{eff}}$ (in the \textit{i}-band) above 5$\arcsec$. Only a handful of the unrecovered objects from NGFS possess an $r_{\mathrm{eff}}$ significantly larger than $5\arcsec$. The NGFS catalog therefore appears to be more complete for objects near our limit of $r_{\mathrm{eff}}=5\arcsec$ and fainter than what volunteers were expected to classify, which is likely a result of the NGFS images being inspected by a team of researchers with higher expertise. From P23, Blurs recovers 397 of the 416 objects (95.4\%) without restricting $r_{\mathrm{eff}}$ to $\ge 5\arcsec$, missing only 19. Some of the unrecovered objects resemble proper DG candidates but appear very faint in surface brightness, likely too faint to be detected by our search. Conversely, Blurs found 167 and 246 objects that NGFS and P23 did not find, but without consistent $r$-band parameters it is unclear if there is any area in the explored parameter space where our citizen science methodology achieves enhanced sensitivity over expert identification. 

\section{Discussion}{\label{sec:discussion}}
\subsection{Citizen Science Performance \label{subsec:citsci_perf}}

 \subsubsection{Recovery Success Across Parameter Space \label{discussion:recovery_success}}
In the parameter space explored by our search—namely DGs with $r_{\mathrm{eff}} \ge 5\arcsec$—the percentage of objects we recover from the majority of Fornax dwarf catalogs, both automated and visual, exceeds 80\%. We achieve $>95\%$ agreement with the full catalogs of SMUDGes, P23, and FCC (dE only), even when including objects smaller than $5\arcsec$. We recover 88.4\%, 80.6\%, and 65.2\% of the V18, NGFS, and V22 catalogs, respectively, when limited to objects with $r_{\mathrm{eff}} \ge 5\arcsec$. V22 shares the least amount of overlap with our parameter space, with only a quarter of the full catalog possessing $r_{\mathrm{eff}} \ge 5\arcsec$. Our lower recovery fraction with V22 thus does not indicate any concerning completeness gap in $r_{\mathrm{eff}}$. The objects we do miss from V22 above $5\arcsec$ and from SMUDGes are near or fainter than $\langle \mu_r \rangle \sim$ 26 mag arcsec$^{-2}$, as shown in Fig. \ref{fig:rMag_reff}. Our catalog moreover sparsely populates the parameter space fainter than this surface brightness, suggesting that this may be the limit at which DGs can be detected by-eye in the images we provided. We observe similarly reduced completeness in the small/faint regime compared to the visual catalogs (NGFS and P23), outside of which there are no concerning completeness gaps that would suggest our citizen science method cannot perform as well as visual inspection by expert researchers. We therefore conclude that for $\langle \mu_r \rangle \lesssim$ 26 mag/arcsec$^{-2}$ and $r_{\mathrm{eff}} \ge 5\arcsec$, our final catalog is highly complete. 

 \begin{figure*}[ht!]
    \centering
    \includegraphics[width=0.75\linewidth]{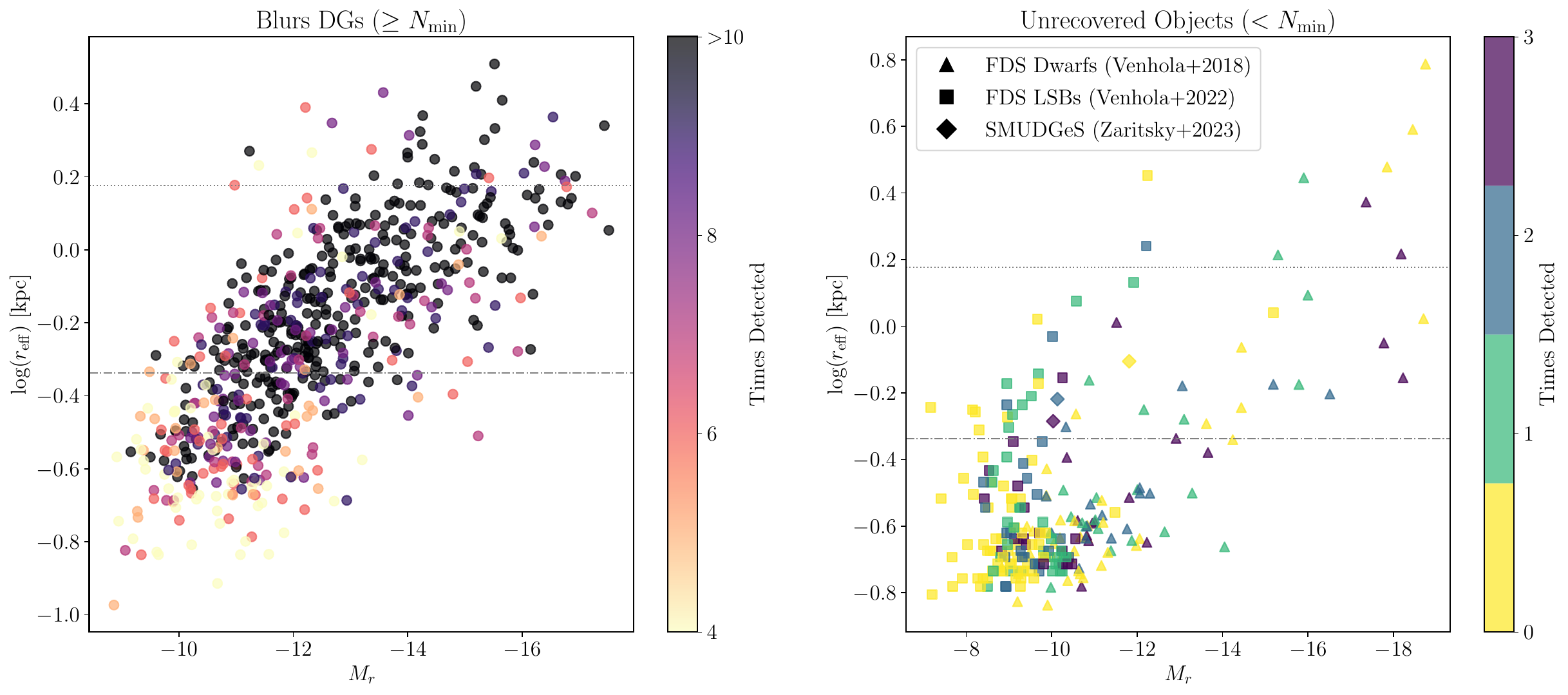}
    \caption{
    Distribution of Zooniverse classifications (point colors) for DGs in Fornax shown in the size--luminosity plane. \textit{Left}: All DGs in the Blurs catalog. \textit{Right}: Known objects in Fornax (from V18, V22, and SMUDGes; represented with the same symbols as in Fig.~\ref{fig:rMag_reff}) that did not meet the threshold number of volunteer detections required to be included in the final Blurs catalog ($N_\mathrm{min} \geq 4$). In the left panel the most often identified objects fill most of the parameter space where we expected volunteers to comfortably classify DGs, and detection counts fall off for fainter and smaller objects. In the right panel the count distribution appears more random, except for the faintest objects which were mostly missed by all volunteers. As done in Figures \ref{fig:rMag_reff} and \ref{fig:ext_coverage}, the horizontal dotted and dash-dotted lines correspond to an $r_{\mathrm{eff}}$ of 1.5 kpc (UDG threshold) and 5\arcsec.}
    \label{fig:count_distribution}
\end{figure*}

Our recovery success is additionally affected by our choice of $N_{\mathrm{min}}=4$, as there may be objects that we did in fact recover from the above catalogs but were ultimately cut from our sample because of a low detection count. Indeed, the detection frequency of our candidates does near or equal $N_{\mathrm{min}}$ in the small and faint regime of the parameter space, as displayed in the left panel of Figure \ref{fig:count_distribution}. However, we show in the right panel of Figure \ref{fig:count_distribution} that although the majority of objects we fail to recover from the automated catalogs are small and faint, they generally exhibit a random scatter in detection frequency below $N_{\mathrm{min}}$. Lowering $N_{\mathrm{min}}$ from 4 would thus not recover a large portion of the parameter space and would introduce hundreds to thousands more false positives to the candidate pool. Thus, $N_{\mathrm{min}}$ of 4 appears to be an appropriate trade-off point between completeness and false positives, and we would likely adopt the same threshold in any future searches. Furthermore, the objects which were detected at least the same number of times as the retirement limit of 10 span the parameter space where we would expect volunteers to comfortably identify DGs, therefore we conclude that this retirement limit is sufficient.

 \subsubsection{Objects in Blurs Missed by Automated Searches \label{discussion:automated_missed}}
In total, our final catalog contains 97 DG candidates not detected by any of the automated searches conducted by V18, V22, or SMUDGes, which we provide DECaLS images of in Appendix \ref{appendix:weFound_NotAutomated}. We propose that V18 and SMUDGes likely missed these objects because of the presence of nearby bright stars and galaxies, making it more difficult for their automated algorithms to accurately detect faint objects, or possibly because these DGs were located under their bright pixel masking. A more detailed study on why these DGs were missed by these automated methods is left for future work. Given that we achieve close agreement with V18 and SMUDGes for $r_{\mathrm{eff}} \ge 5\arcsec$, we subsequently exceed the completeness of V18 and SMUDGes within their intended parameter space based on candidate totals alone, though a more rigorous recovery analysis with injected sources (as done by SMUDGes) would further validate our completeness estimates. In a future citizen science search, similar artificial objects could be injected into a subset of the images before they are inspected by the volunteers, as a means of producing an equivalent completeness estimate. We are unable to identify any surprising candidates from Blurs missed by V22 due to V22 having a much narrower sensitivity space than V18 and a majority of objects with $r_{\mathrm{eff}} < 5\arcsec$, where we are more incomplete.

\subsubsection{Efficiency}
Our set of nearly 15,000 images (spanning 26 deg$^2$) was fully inspected in roughly 30 days, due to the participation of over 1,400 Zooniverse volunteers, corresponding to an average inspection rate of 1.15 days per square degree. In \cite{Dey_2025}, the same Zooniverse approach as this work was utilized to examine $\sim$100 deg$^2$ of the Virgo cluster and was completed in roughly 140 days (1.4 days/deg$^2$), which is comparable to the Fornax inspection rate. We estimate that this method would exceed a completion time of one year once the targeted sky area exceeds $\sim$300 deg$^2$, assuming a similar level of public interest as our Fornax and Virgo searches. If a single person wanted to look through the entire Fornax image set in 30 days time, one would have to look at approximately 500 images every day. Although this may be practical for some, our citizen science approach produces a complete catalog of DGs in a reasonable amount of time by distributing this workload over a large number of interested, non-expert volunteers. We observed that volunteers who posted to the project discussion boards expressed little to no difficulty in comprehending the classification workflow, thus indicating that the proposed tasks are adequately accessible and can be completed efficiently. 
 
If a project of this kind were to receive volunteer participation on the order of tens of thousands of people, as seen with projects such as \textit{Galaxy Zoo} \citep{Lintott_2008} and \textit{Backyard Worlds} \citep{BackyardWorlds_2017}, an all-sky visual search for DGs in all environments would become feasible. This could be achieved with the upcoming Legacy Survey of Space and Time (LSST), which will survey 18,000 square degrees of the sky with the best optical depth to date for identifying low surface brightness objects like DGs \citep{LSST_2019}. Approximately 10\% of LSST time will be dedicated to observing several smaller sky areas with deeper coverage, known as Deep Drilling Fields, which would be achievable to examine with our methodology even with the same level of participation as our Fornax search. Other potential directions for this work include exclusively searching for DGs near bright stars —where we do better than automated methods—to increase completeness (e.g., across 100 deg$^2$ or more), potentially with the Legacy Survey or in the future with LSST. 

\subsubsection{Classification Accuracy of Nucleated Dwarfs}
In Figure \ref{fig:nucl_accuracy}, we explore how reliably the volunteers distinguish between nucleated and non-nucleated diffuse galaxies in our final catalog. We consider the volunteer classification of an object the most voted response (either nucleated or non-nucleated) according to some minimum level of majority consensus ($> 50\%$ of votes) and compare this answer to our expert classification. At the weakest consensus thresholds ($\sim$50-60\%), the volunteers identify non-nucleated objects with $>$97.5\% accuracy and nucleated objects with $>$82.5\% accuracy. Increasing the consensus threshold further improves classification accuracy, but also reduces the available sample size, as fewer objects will have classifications counts that meet the required threshold. For a consensus threshold of $\geq$84\%, volunteer classifications become identical to expert classifications. These performance metrics indicate that most volunteers can capably recognize a DG as non-nucleated but may struggle more with labeling a DG as nucleated, which is to be expected given that an NSC will not always be bright, centered, or isolated. In future citizen science searches for DGs, imposing a consensus threshold of at least 84\% may be useful for reducing the amount of candidates requiring expert inspection to be formally classified as nucleated or non-nucleated. In addition, even with a much lower threshold, (non-)nucleated classifications from volunteers remain overwhelmingly accurate.

\begin{figure}[ht!]
    \centering
    \includegraphics[width=1.0\linewidth]{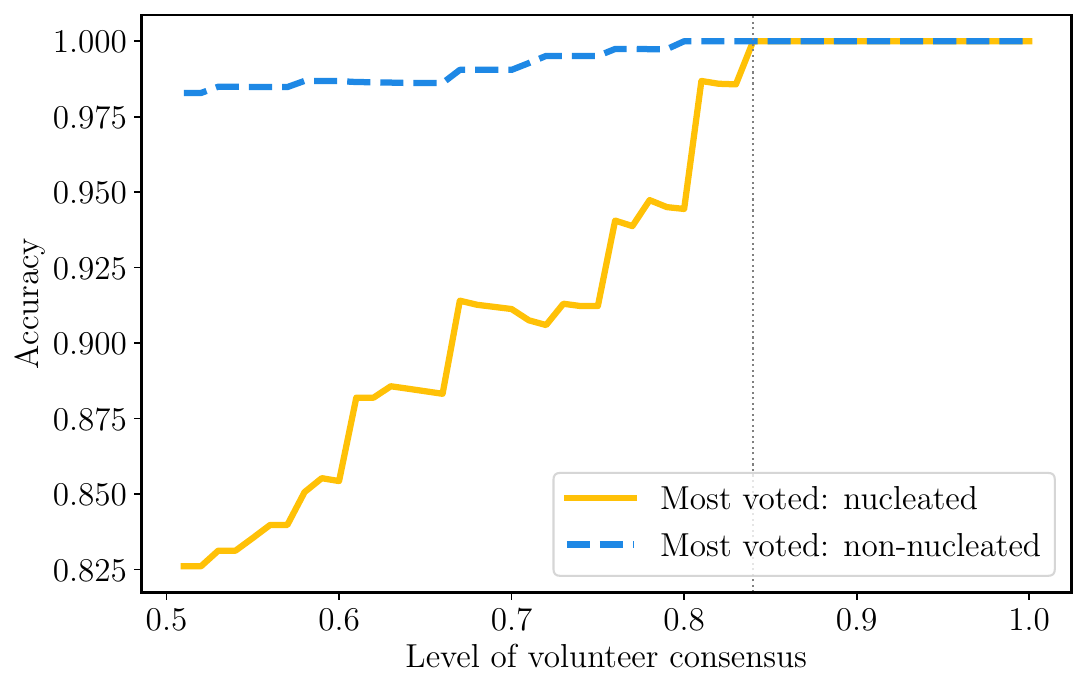}    \caption{Accuracy of volunteers when distinguishing between nucleated and non-nucleated diffuse galaxies according to the strength of the volunteer consensus, i.e., the most voted response by a $>$50\% majority selecting either `nucleated' or `non-nucleated'. We compare the volunteer-based classification of the objects in our final catalog to our expert classification to compute the volunteer accuracy over a range of consensus levels from 50\% to 100\%. At a consensus strength of 84\% (indicated by dotted vertical line), the volunteer classification of both nucleated and non-nucleated objects becomes identical to our expert classification.}
    \label{fig:nucl_accuracy}
\end{figure}

\subsection{Nucleated Population}
In this subsection, we examine the spatial distribution, mass, and NSC-to-galaxy light fraction of the nucleated galaxies in our catalog to better understand the origins of nuclear star clusters and diffuse galaxies themselves. Our main findings are the following: we observe an apparent decrease in the density of non-nucleated DGs near the cluster center that may hint at a biased formation of NSCs driven by globular cluster inspiral (Section \ref{discussion:radial_distr}). Secondly, we find that the nucleated fraction peaks at over 80\% for a host galaxy stellar mass of $\sim10^{8.5} M_{\odot}$, which is consistent with dwarf studies of Fornax and Virgo in the literature (Section \ref{discussion:stellar_mass}). Finally, we find the NSC mass distribution of our catalog to be approximately Gaussian with a mean mass of $10^{5.773} M_{\odot}$, characterized by a typical NSC-to-galaxy light fraction between 0.1 to 5\%, though we highlight four cases where it exceeds 10\% (Section \ref{discussion:luminosity_ratio}). We discuss these results as they pertain to possible NSC formation pathways in the subsequent paragraphs. 

\subsubsection{Current Models for DG \& NSC Formation}
The origins of diffuse galaxies remains poorly constrained. It has been theorized that UDGs formed in Milky Way-sized dark matter halos but then fell into the cluster environment early on in their evolution and had their growth stunted by an extreme level of environmental feedback \citep{VanDokkum_2016}. Another proposed model is that UDGs come from dwarf galaxies that have been expanded due to strong feedback, which then retain their enlarged size as they fall into the cluster environment (\citealp{DiCintio_2017}; \citealp{Chan_2018}). Interactions with other galaxies or the cluster potential may similarly transform normal dwarf galaxies into UDGs through tidal stripping and heating (\citealp{Bennet_2018}; \citealp{Carleton_2019}; \citealp{Jones_2021_UDGs}; \citealp{Fielder_2024}). These mechanisms may also be responsible for the formation of sub-UDG scale diffuse galaxies. The formation scenario of the host galaxy is further complicated by the presence of a nuclear star cluster. Two leading proposed mechanisms for the formation of NSCs are globular cluster infall caused by dynamical friction \citep{Tremaine_1975,Lotz_2001} and in situ star formation as a result of gas funneled into the galaxy center \citep{VanDenBergh_1986}. Incorporating dissipative and feedback-regulated processes, such as supernova outflows and dynamical heating from a central massive black hole, represent ongoing efforts to construct a more complete picture of NSC growth  (\citealp{McLaughlin_2006}; \citealp{Antonini_2015}).

\subsubsection{Radial Distribution \label{discussion:radial_distr}}

\begin{figure*}[ht!]
    \centering
    \includegraphics[width=0.75\linewidth]{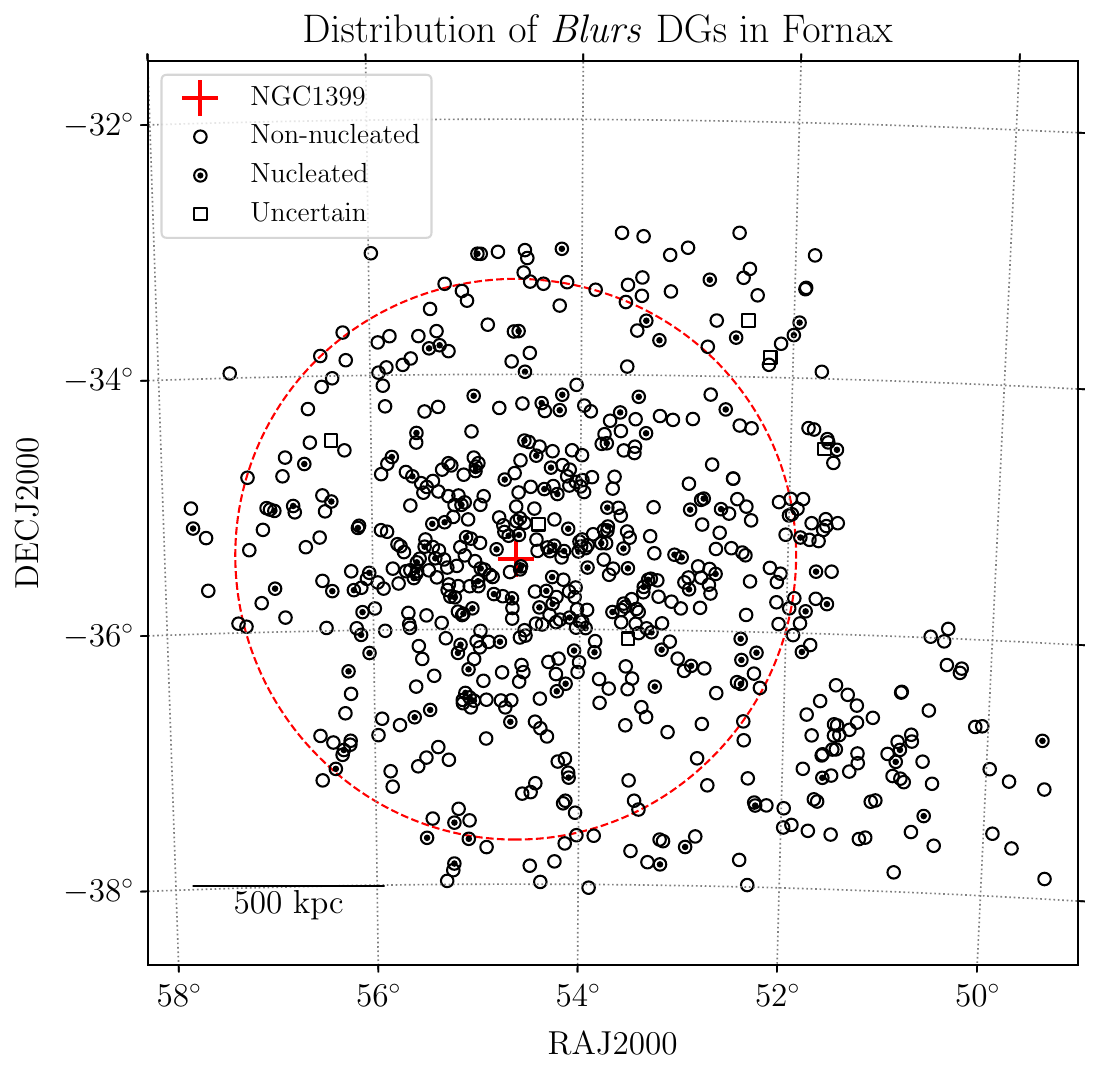}
    \caption{Spatial distribution in Fornax of diffuse galaxy candidates from our final catalog. Non-nucleated DGs are indicated by the open black circles, while nucleated DGs have an additional central black dot. The red dashed circle denotes the Fornax cluster virial radius, $R_{\mathrm{vir}}$, of 0.7 Mpc (2.2$^{\circ}$), centered on the central cluster galaxy, NGC1399, represented by the red cross.}
    \label{fig:cands_sky_plot}
\end{figure*}

The spatial distribution in Fornax of the nucleated and non-nucleated candidates is presented in Figure \ref{fig:cands_sky_plot}. Approximately 70\% of objects reside within the Fornax virial radius of 0.7 Mpc, with the furthest object being 1.6 Mpc from the central cluster galaxy, NGC 1399. We report a total of 139 nucleated and 498 non-nucleated diffuse galaxies, yielding a global nucleated fraction of 21.8\%. In six cases, we were unable to visually classify an object as nucleated or non-nucleated for reasons such as an exceptionally faint NSC-candidate, the NSC-candidate being significantly offset from the galaxy center, or multiple NSC-like structures. We label these DGs as ``uncertain" in our final catalog and exclude them from our calculation of the nucleated fraction (displayed in Appendix \ref{appendix:uncertain}). We find that the nucleated fraction reaches 25\% within $R_\mathrm{vir}$ and 35\% within the Fornax core region ($R_\mathrm{vir}/4$).

\begin{figure*}[ht!]
    \centering
    \includegraphics[width=0.75\linewidth]{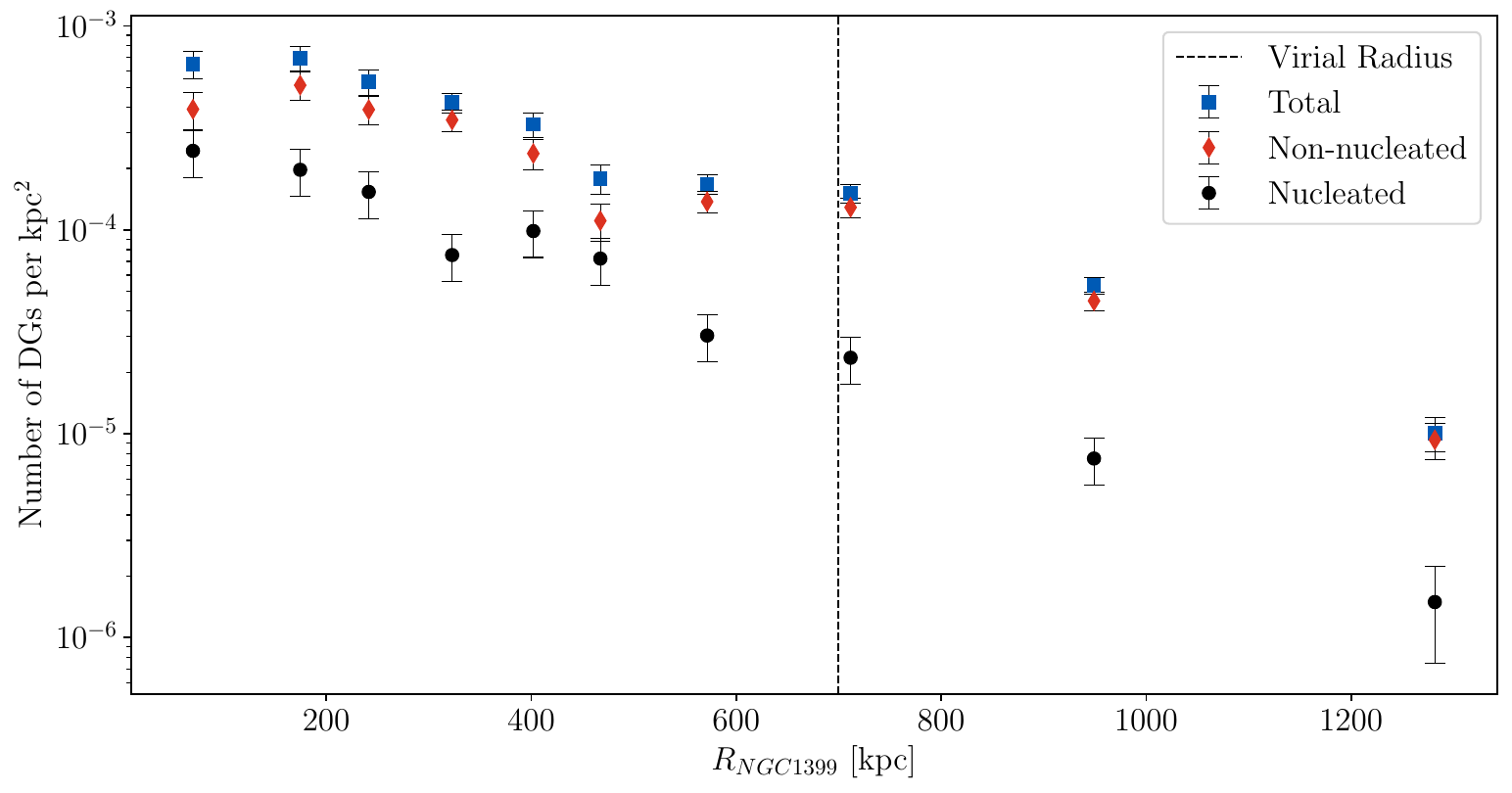}
    \caption{Total, non-nucleated, and nucleated number density profiles of our Blurs catalog as a function of radial distance from NGC 1399, the galaxy considered to be at the center of the Fornax cluster. The vertical dashed line denotes the Fornax virial radius of 700 kpc. At $R_{NGC1399} >$ 700 kpc, the values become less reliable because of asymmetries in the cluster structure. A possible decrease in the non-nucleated density is observed near the cluster center. Around 400 kpc, an apparent spike in the nucleated density is also observed. }
    \label{fig:radial_density}
\end{figure*}

In Figure \ref{fig:radial_density}, we plot the radial number density distribution of the nucleated, non-nucleated, and all (including uncertain class) objects from our catalog. Due to the lower overall number of nucleated DGs than non-nucleated ones, we bin the nucleated objects radially in consecutive groups of 15 then use the radial span of each bin as the annuli for computing the non-nucleated and total densities. These annuli are on the order of $\sim$50-100 kpc in width except beyond $R_{\mathrm{vir}}$, where the width exceeds 200 kpc. We find the number density of both nucleated and non-nucleated DGs generally decreases with increasing radial distance from the cluster center, though we observe a possible decline in the non-nucleated density within the central 200 kpc. We note that the number densities become less reliable past $R_\mathrm{vir}$ because of the asymmetry in the cluster shape created from a sub-cluster around NGC 1316 falling into the main Fornax cluster \citep{Drinkwater_2001}. If real, a decrease in non-nucleated DGs near the center alongside an increase in the nucleated density could be attributed to the increased rate of galaxy harassment in the cluster core. The higher concentration of galaxies in the core would induce more frequent perturbations of DGs than in the cluster outskirts which may promote globular cluster inspiral in initially non-nucleated DGs with multiple GCs. We also observe an apparent spike in the density of nucleated DGs around 400 kpc. This rise is seen in two adjacent bins and appears to be a real feature of the distribution. There are no gaps in the FDS coverage near 400 kpc that would explain this increase, nor any obvious structures in the cluster at this distance that could offer an explanation for this feature, therefore more investigation is needed. 

\subsubsection{Stellar Mass Distribution \label{discussion:stellar_mass}}

 \begin{figure}[ht!]
    \centering
    \includegraphics[width=1.0\linewidth]{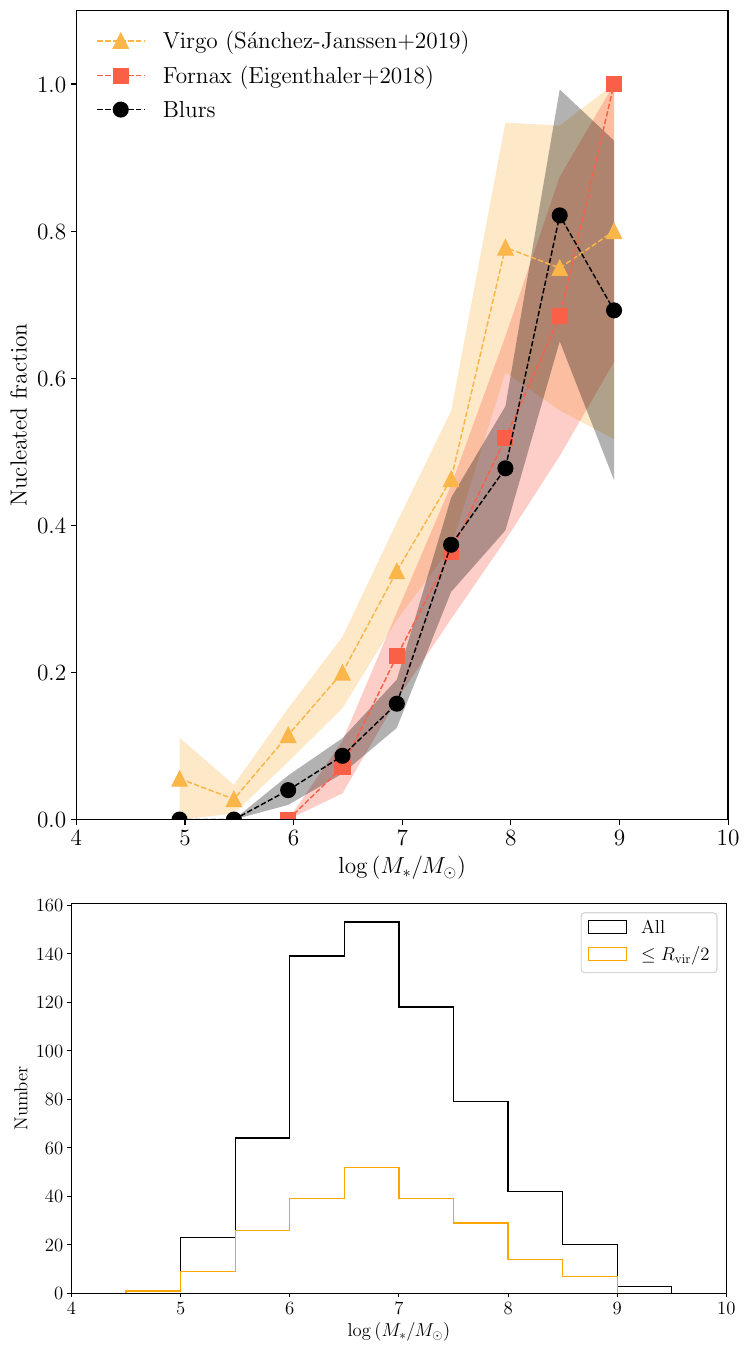}
    \caption{\textit{Upper:} Nucleated fraction of our final catalog (black dots) as a function of stellar mass, alongside other studies of Fornax (red squares; \citealp{Eigenthaler_2018}) and the Virgo cluster (orange triangles; \citealp{Sanchez-Janssen_2019}). We find that our results agree well with the Fornax curve but the curve for Virgo has a higher nucleated fraction relative to Fornax (cf. figure 2 of \citealp{Sanchez-Janssen_2019}). \textit{Lower:} Stellar mass distribution of the DGs in our final catalog for the entire Fornax cluster (black) and within half of the cluster virial radius (orange).}
    \label{fig:stellar_mass}
\end{figure}

\begin{figure*}[ht!]
    \centering
    \includegraphics[width=0.75\linewidth]{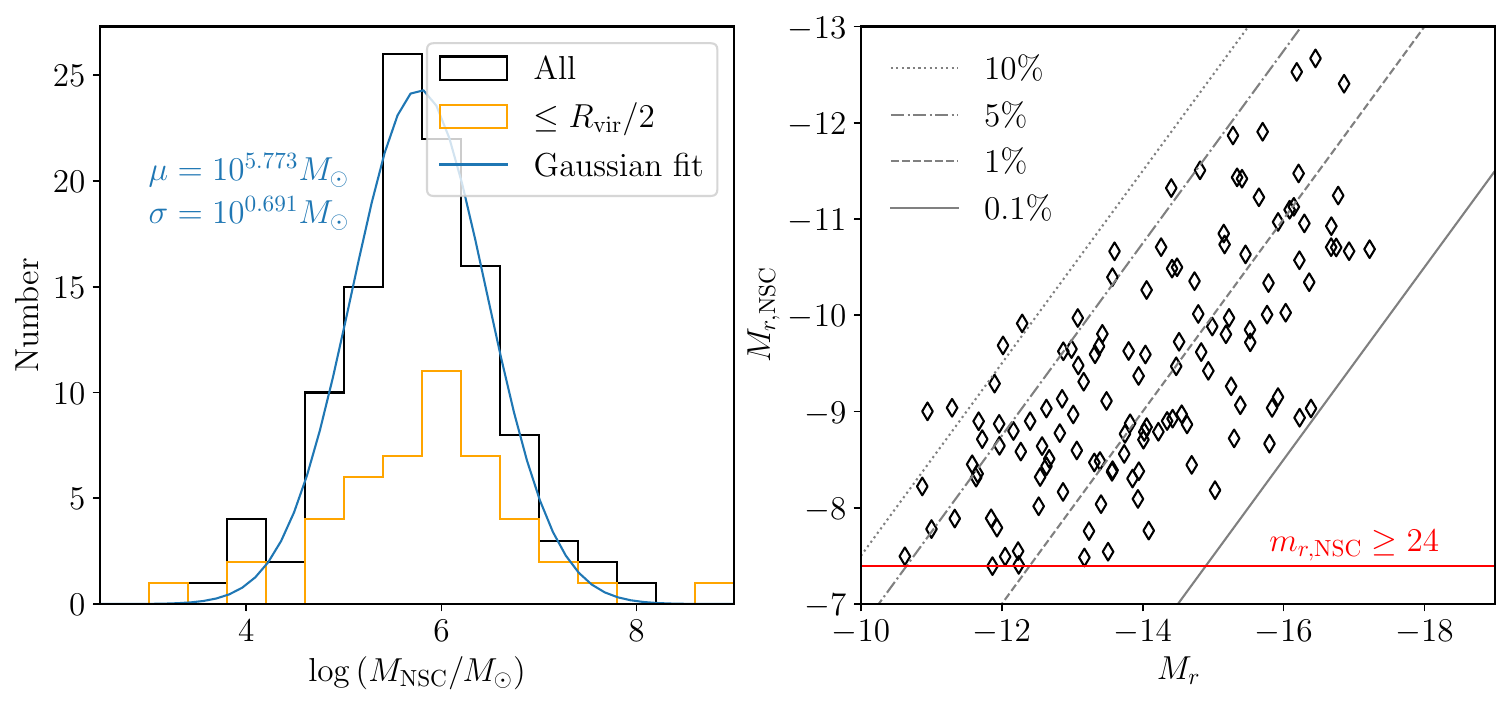}
    \caption{\textit{Left:} Stellar mass distribution of the nuclear star clusters in our final catalog (113 nuclei) plotted for the entire Fornax cluster (black line) and within half the cluster virial radius (orange line). The cluster-wide distribution is approximately Gaussian (blue line) with a mean NSC mass $M_{\mathrm{NSC}}=10^{5.773}M_{\odot}$ and standard deviation $\sigma=10^{0.691}M_{\odot}$. \textit{Right:} Absolute \textit{r}-band magnitude of each NSC in our final catalog ($M_{r, \mathrm{NSC}}$) versus the absolute \textit{r}-band magnitude of its corresponding host galaxy ($M_r$). The diagonal lines denote NSC-to-host galaxy luminosity ratios of 10\%, 5\%, 1\%, and 0.1\%, from left to right. Four candidates notably exceed a luminosity ratio of 10\% (see Fig. \ref{fig:nsc_over10}). NSCs with an apparent \textit{r}-band magnitude fainter than 24 represent the NSC detection limit of our catalog (horizontal red line).}
    \label{fig:nsc_mass_light}
\end{figure*}

We estimate the stellar masses of the DGs in our catalog using the effective stellar mass-to-light ratio ($M_{*}/L$) as a function of \textit{g-r} color given by  \cite{Zibetti_2009}. Converting from $M_{*}/L$ to galaxy stellar mass $M_*$ in solar units, we incorporate the absolute solar magnitude \citep{Willmer_2018} for the SDSS\_r filter (most similar to FDS) into the \cite{Zibetti_2009} equation as follows:

\begin{equation}\label{eq:stellar_mass}
\begin{aligned}
    \log(M_{\ast}/M_{\odot}) = & -0.840 + 1.654(g-r) \\
    & - 0.4(M_{r,\ast} - M_{r,\odot})
\end{aligned}
\end{equation}

\noindent where $M_{r,\ast}$ is the absolute magnitude of the host diffuse galaxy in the \textit{r}-band, $M_{r,\odot}$ is the absolute solar magnitude (4.53 mag), and -0.840 and 1.654 are the \textit{g-r} calibration parameters provided \cite{Zibetti_2009}. 

In the top panel of Figure \ref{fig:stellar_mass}, we aim to reproduce figure 2 from \cite{Sanchez-Janssen_2019} to investigate the nucleated fraction of our catalog as a function of host galaxy stellar mass. The study conducted by \cite{Sanchez-Janssen_2019} focuses on quiescent galaxies in the Virgo cluster core with data from the Next Generation Virgo Survey (NGVS). We include these data in Fig. \ref{fig:stellar_mass} to compare our distribution to another cluster environment, as well as data from an NGFS study of dwarf galaxies in the Fornax core region \citep{Eigenthaler_2018}. Our resulting mass distribution agrees well with the distribution found in Fornax by \cite{Eigenthaler_2018}, peaking at a nucleated fraction of 80\% near a galaxy stellar mass of $\sim$10$^{8.5}M_{\odot}$. However, we find that the nucleated fraction in Virgo appears to be higher than Fornax for almost all $M_*$, whereas in \cite{Sanchez-Janssen_2019} these two distributions exhibit nearly identical behavior. It is unclear where this deviation in our plot is coming from, since we rely on the stellar masses reported in the catalogs by \cite{Sanchez-Janssen_2019} and \cite{Eigenthaler_2018} rather than recomputing $M_*$ ourselves according to Eq. \ref{eq:stellar_mass}. We propose this could possibly be due to a bias in stellar mass calculation, as \cite{Eigenthaler_2018} follows the same parameterization of $M_{*}/L$ to calculate $M_*$ but with the color calibrations defined by \cite{Bell_2003}, while \cite{Sanchez-Janssen_2019} employs a different approach that involves modeling the spectral energy distributions in the \textit{u'griz'} bands. We provide a histogram of our $M_*$ distribution for the entire cluster and within $R_{\mathrm{vir}}/2$ in the bottom panel of Figure \ref{fig:stellar_mass}, peaking at $\sim$10$^{7}M_{\odot}$ for both populations. 

We compute the NSC masses ($M_{\mathrm{NSC}}$) for 113 nuclei (out of 139) from our final catalog using the same method defined previously for $M_*$. Robust photometric parameters for the remaining 26 NSCs could not be obtained with GALFIT due to poor fitting conditions caused by nearby bright objects, therefore we exclude these 26 nuclei from our analysis. The nucleated dwarfs from \cite{Eigenthaler_2018} in the inner cluster region ($R_{\mathrm{NGC1399}} \lesssim 350$ kpc) were used in \cite{Ordenes-Briceno_2018_NSC} to study their nuclei, from which they found a bimodal NSC mass distribution that peaks at $\simeq 10^{5.4} M_{\odot}$ and has a second peak at $10^{6.3} M_{\odot}$. In contrast, our $M_{\mathrm{NSC}}$ distribution (left panel of Figure \ref{fig:nsc_mass_light}) does not show any clear signs of bimodality for either the total nucleated population or the objects within 350 kpc of the cluster center ($R_{\mathrm{vir}}/2$). Instead, the distribution for all NSCs is approximately Gaussian with $\langle M_{\mathrm{NSC}} \rangle= 10^{5.773} M_{\odot}$ and a standard deviation of 0.691 dex. The origin of this discrepancy with \cite{Ordenes-Briceno_2018_NSC} is unclear. We note that the bimodality in NSC mass may be linked to the mild bimodality they observe in the corresponding dwarf spheroid mass distribution, as these two quantities are correlated (they find that $M_{\mathrm{NSC}}/M_{*} \propto M_*^{-0.5}$); however, \cite{Ordenes-Briceno_2018_NSC} ultimately concluded that this bimodality in $M_*$ was statistically insignificant. The NSC masses reported by \cite{Ordenes-Briceno_2018_NSC} were also computed through a different mass estimation technique involving a $\chi^2$ minimization approach based on several filters not included in the FDS photometric coverage.

\subsubsection{NSC-to-Host Galaxy Luminosity Ratio \label{discussion:luminosity_ratio}}
We use the NSC photometry in the \textit{r}-band to compute the NSC-to-host galaxy luminosity ratio, i.e., the fraction of the total galaxy light emitted by the NSC, to probe the NSC evolutionary status. Higher ratios have been found to suggest ongoing in situ star formation as the predominant mechanism for NSC growth, while lower ratios appear more consistent with an NSC growth modeled by globular cluster inspiral \citep{denBrok_2014}. We plot the absolute NSC magnitude in \textit{r} ($M_{r,\mathrm{NSC}}$) versus the host galaxy absolute \textit{r}-band magnitude ($M_r$) in the right panel of Figure \ref{fig:nsc_mass_light}. We find that nearly all of the nucleated DGs in our final catalog exist between 0.1\% and 5\% fractional luminosity, which is expected given our NSC mass distribution is centered around $10^{5.773}M_{\odot}$. Four candidates from our catalog present with fractions over 10\%, which we display in Figure \ref{fig:nsc_over10}. Of the four candidates, BLUR033632-362539 and BLUR034747-353936 have an estimated NSC mass of $10^{6.14}M_\odot$ and $10^{6.59}M_\odot$, respectively, similar to low mass ultracompact dwarf candidates. The vicinity around each candidate was examined in the Legacy Survey Sky Viewer to check for possible tidal stripping from nearby galaxies, however, only BLUR033956-353723 was close enough for an associated galaxy to be identified. This candidate is $\sim$13.85 kpc from NGC 1427A, the brightest irregular galaxy in the Fornax cluster, but we do not see any evidence of a tidal stream between the two objects in the Legacy Survey. Interestingly, the range of fractional luminosities we observe for our nucleated sample (0.1-10\%) is similar to the range found by \cite{Khim_2024} for the fraction of stellar mass in the NSC. 

\begin{figure*}[ht!]
    \centering
    \includegraphics[width=0.75\linewidth]{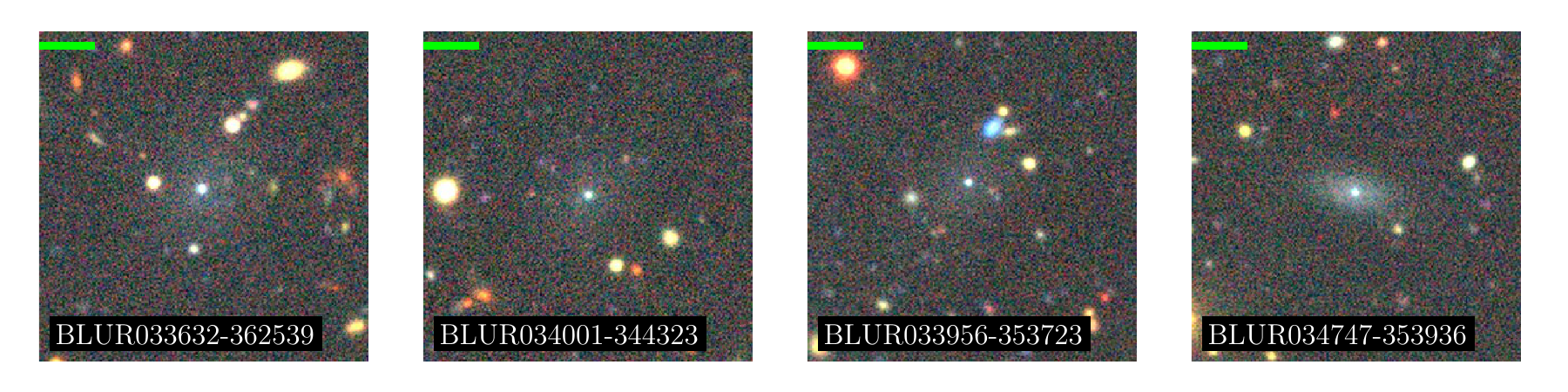}
    \caption{DECaLS images of four candidates from Blurs that possess an NSC-to-host galaxy luminosity ratio (in the \textit{r}-band) above 10\%. These objects are of interest for understanding low mass ultracompact dwarfs (UCDs) that may have formed through tidal stripping from nearby galaxies. No such galaxies were found within the vicinity of each object except for BLUR033956-353723, though no evidence of a tidal stream was found in the Legacy Viewer. The green bar in each image denotes 10\arcsec in the image scale.}
    \label{fig:nsc_over10}
\end{figure*}

\section{Summary and Conclusions}{\label{sec:concl}}
We present a catalog of 643 diffuse galaxies (DGs) in the Fornax cluster identified with our citizen science project, ``Blobs and Blurs: Extreme Galaxies in Clusters", across 26 square degrees of the Fornax cluster. Over 1,400 volunteers participated in this project on Zooniverse, helping us to examine nearly 15,000 image cutouts of optical data obtained from the Fornax Deep Survey (FDS) and Dark Energy Camera Legacy Survey (DECaLS). This marks the first effort to produce a cluster-wide catalog of diffuse galaxies strictly with citizen science. Our key results are summarized as follows: 

(i) Our catalog is highly complete relative to nearly all dwarf catalogs in Fornax, within the parameter space targeted by our search, namely objects with $r_{\mathrm{eff}} \ge 5\arcsec$ and $\langle \mu_r \rangle \lesssim$ 26 mag arcsec$^{-2}$. These catalogs consist of both automated searches (V18, \citealp{Venhola_2018}; V22, \citealp{Venhola_2022}; SMUDGes, \citealp{Zaritsky_2023}) and visual ones (NGFS, \citealp{Munoz_2015}, \citealp{Eigenthaler_2018}, \citealp{Ordenes-Briceno_2018}; P23, \citealp{Paudel_2023}; FCC, \citealp{Ferguson_1989}). We recover $>80\%$ of the objects from the following catalogs: SMUDGes, P23, FCC (dE galaxies only), as well as V18 and NGFS when limited to $r_{\mathrm{eff}} \ge 5\arcsec$. Our recovery fraction is lowest with V22, recovering only 65.2\% of objects even when restricted to $r_{\mathrm{eff}} > 5\arcsec$. This is likely due to the fact that V22 was optimized to find LSB galaxies fainter than what could realistically be found by-eye in the images we provided, i.e., fainter than $\langle \mu_r \rangle \simeq$ 26 mag arcsec$^{-2}$. The few objects we miss from SMUDGes are similarly positioned near this same surface brightness, thus this appears to be a real limitation on our completeness.  

(ii) Our catalog identifies 97 DG candidates that the automated detection algorithms employed by V18, V22, and SMUDGes collectively did not find. Of these 97 objects, 26 are in the range we would expect V18 and SMUDGes to be sensitive to ($r_{\mathrm{eff}} \ge 5\arcsec$). We propose that these objects were likely not detected due to the presence of a nearby bright star or galaxy, thus rendering our catalog more complete under these conditions. We additionally identify 3 diffuse galaxy candidates in Fornax not detected by any prior searches, automated or visual, that also do not appear in any extragalactic database. These galaxies are near the minimum size limit of our search, therefore observations of their recession velocities are needed to determine if they are members of the Fornax cluster and not background galaxies. 

(iii) Our search was completed in $\sim$30 days, equating to an average classification rate of 1.15 days per square degree. Assuming a similar level of participation in future searches, our citizen science approach would thus be most practical for examining sky surveys on the order of $\sim$100 to 300 square degrees. However, with 10 to 100 times more volunteers, an all-sky search for DGs could become a feasible goal for future applications of this project, such as with deeper optical imaging from LSST.  

(iv) Volunteers demonstrate an ability to reliably distinguish between nucleated and non-nucleated diffuse galaxies. We find that when at least 84\% of volunteers agree on the same classification of an object (nucleated or non-nucleated), they are able to identify nucleated and non-nucleated objects in our final catalog 100\% accurately. This metric may prove useful in future citizen science searches for DGs to minimize the amount of expert oversight needed to produce an accurate catalog. Futhermore, even at much lower consensus thresholds, volunteers were able to accurately classify (non-)nucleated dwarfs the vast majority of the time.

(v) We report a total of 139 nucleated objects in our final catalog, 498 non-nucleated objects, and 6 candidates with uncertain nucleated status, resulting in an overall nucleated fraction of 21.8\% (excluding uncertain class). Within the cluster virial radius, we estimate the nucleated fraction to be $\sim$25\%, which is similar to current estimates in the literature. We find that the number density of nucleated objects in general decreases with distance from the cluster center, except at around 400 kpc, where an apparent spike is observed. No obvious structures in the cluster, nor any changes in the observational coverage, were found to offer a possible explanation for this feature. The density of non-nucleated DGs also generally decreases with distance but appears to decline near the cluster center. This may be the result of a higher rate of galaxy perturbations in the cluster core which drives the formation of nuclear star clusters via globular cluster inspiral. 

(vi) We find that the nucleated fraction of our catalog peaks at over 80\% for a host galaxy stellar mass of $\sim10^{8.5} M_{\odot}$, which is consistent with similar studies of dwarf galaxies in Fornax and Virgo (\citealp{Eigenthaler_2018}; \citealp{Sanchez-Janssen_2019}). Of our nucleated candidates, we report a mean NSC mass of $10^{5.773 \pm 0.691} M_{\odot}$. We do not observe a bimodal NSC mass distribution within 350 kpc of the cluster center as observed by \cite{Ordenes-Briceno_2018_NSC} and it is unclear why this discrepancy exists.  

Our final catalog is a relatively complete set of diffuse galaxy candidates in the Fornax cluster. We have demonstrated that our citizen science methodology is both reliable and efficient enough to enable the visual construction of large sky area DG catalogs that might otherwise be too time-consuming to inspect with a small team of researchers alone. Furthermore, this approach can outperform individual search algorithms in a number of scenarios, and all that have been used to date in Fornax in the vicinity of bright objects. Although the Zooniverse search has concluded, the project can easily be updated with data from new imaging surveys to accomplish future searches for DGs. 

\section*{Acknowledgments}
The science in this publication was made possible by the participation of over 1,400 volunteers in the \textit{Blobs and Blurs} citizen science project. We gratefully acknowledge their support, as this project would not have been achievable without volunteer engagement. We thank Dennis Zaritsky and Richard Donnerstein for helpful discussions concerning our comparison to SMUDGes sources in Fornax.

This publication uses data generated via the Zooniverse.org platform, development of which is funded by generous support, including from the National Science Foundation, NASA, the Institute of Museum and Library Services, UKRI, a Global Impact Award from Google, and the Alfred P. Sloan Foundation. 

This material is based upon work supported by the National Science Foundation Graduate Research Fellowship Program under Grant No. 24-591. Any opinions, findings, and conclusions or recommendations expressed in this material are those of the author(s) and do not necessarily reflect the views of the National Science Foundation. Supported in part through the Arizona NASA Space Grant Consortium, Cooperative Agreement 80NSSC20M0041. DJS acknowledges support from NSF grant AST-2508746.

Based on data obtained from the ESO Science Archive Facility with DOI(s): \url{https://doi.org/10.18727/archive/23}. This research also relies on images from the Dark Energy Camera Legacy Survey (DECaLS; Proposal ID 2014B-0404; PIs: David Schlegel and Arjun Dey). Full acknowledgment for DECaLS can be found at \url{https://www.legacysurvey.org/acknowledgment/}.

\vspace{5mm}
\facilities{VST, GALEX, Cerro Tololo Inter-American Observatory (CTIO)}

\software{\texttt{astropy} \citep{astropy:2013, astropy:2018, astropy:2022}, 
\texttt{DS9} \citep{SAODS9},
\texttt{GALFIT} \citep{Peng_2010}, 
\texttt{matplotlib} \citep{Hunter:2007}, 
\texttt{numpy} \citep{numpy}, 
\texttt{pandas} \citep{mckinney-proc-scipy-2010, pandas},
\texttt{photutils} \citep{Photutils},  
\texttt{python} \citep{python}, 
\texttt{reproject}\citep{Robitaille},
\texttt{scipy} \citep{2020SciPy-NMeth, scipy}}

\bibliography{main}{}
\bibliographystyle{aasjournal}

\appendix
\section{Galfit Photometry \label{appendix:galfit} }
We assess the quality of our GALFIT photometry by comparing to the GALFIT photometry of V18 and SMUDGes for objects in common with our final catalog. We compare our \textit{r}-band measurements of the galaxy apparent magnitude ($m_r$), NSC apparent magnitude ($m_{r,\mathrm{NSC}}$), and effective radius ($r_{\mathrm{eff}}$) to the estimates obtained by V18 and SMUDGes in Figure \ref{fig:phot_accuracy}, finding a high level of agreement. 

\begin{figure*}[ht!]
    \centering
    \includegraphics[width=0.75\linewidth]{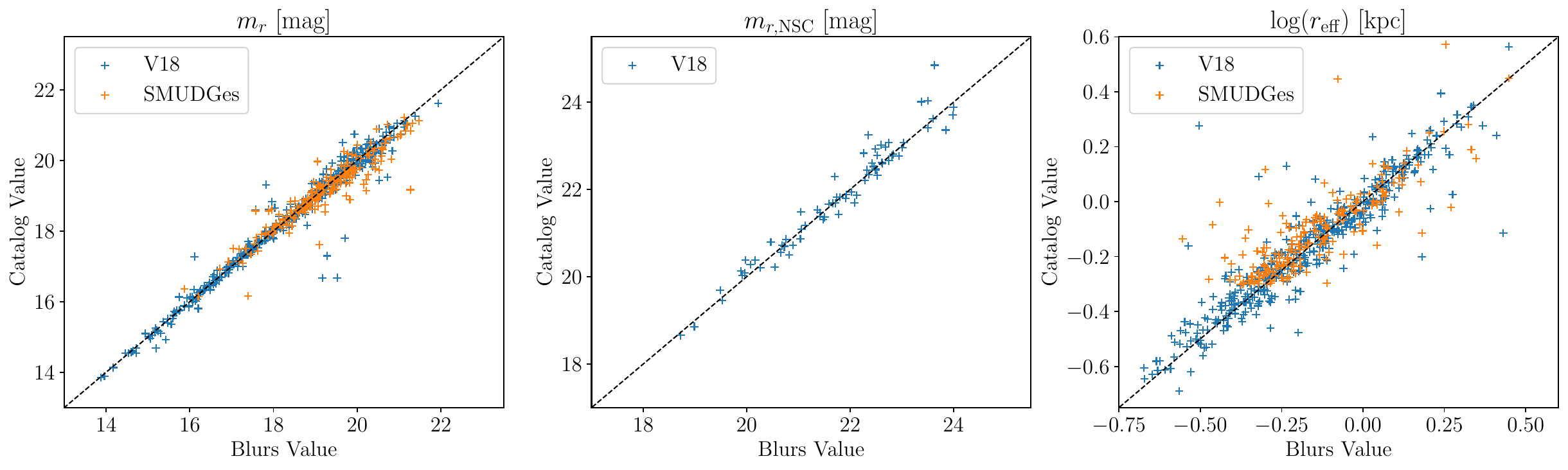}
    \caption{One-to-one plots of our \textit{r}-band GALFIT photometry for $m_r$ (left), $m_{r, \mathrm{NSC}}$, and $r_{\mathrm{eff}}$ (right) with the GALFIT photometry performed by V18 and SMUDGES of the same objects ($m_{r, \mathrm{NSC}}$ not available in SMUDGes). The tight one-to-one relation for these three parameters indicates that our photometry is in good agreement with V18 and SMUDGes, with no significant systematic offset.}
    \label{fig:phot_accuracy}
\end{figure*}

\section{Objects with Uncertain Nucleated Status \label{appendix:uncertain} }
In Figure \ref{fig:uncertain_nucl}, we display DECaLS images for the six candidates from our final catalog that we could not conclusively classify as either nucleated or non-nucleated, and which were thus labeled as ``uncertain". For all of the figures below, a brighter and higher contrast image stretch is used for fainter objects while the default stretch from Legacy Viewer is used for brighter ones. The green bar in each image represents 10\arcsec.

\begin{figure*}[ht!]
    \centering
    \includegraphics[width=0.45\linewidth]{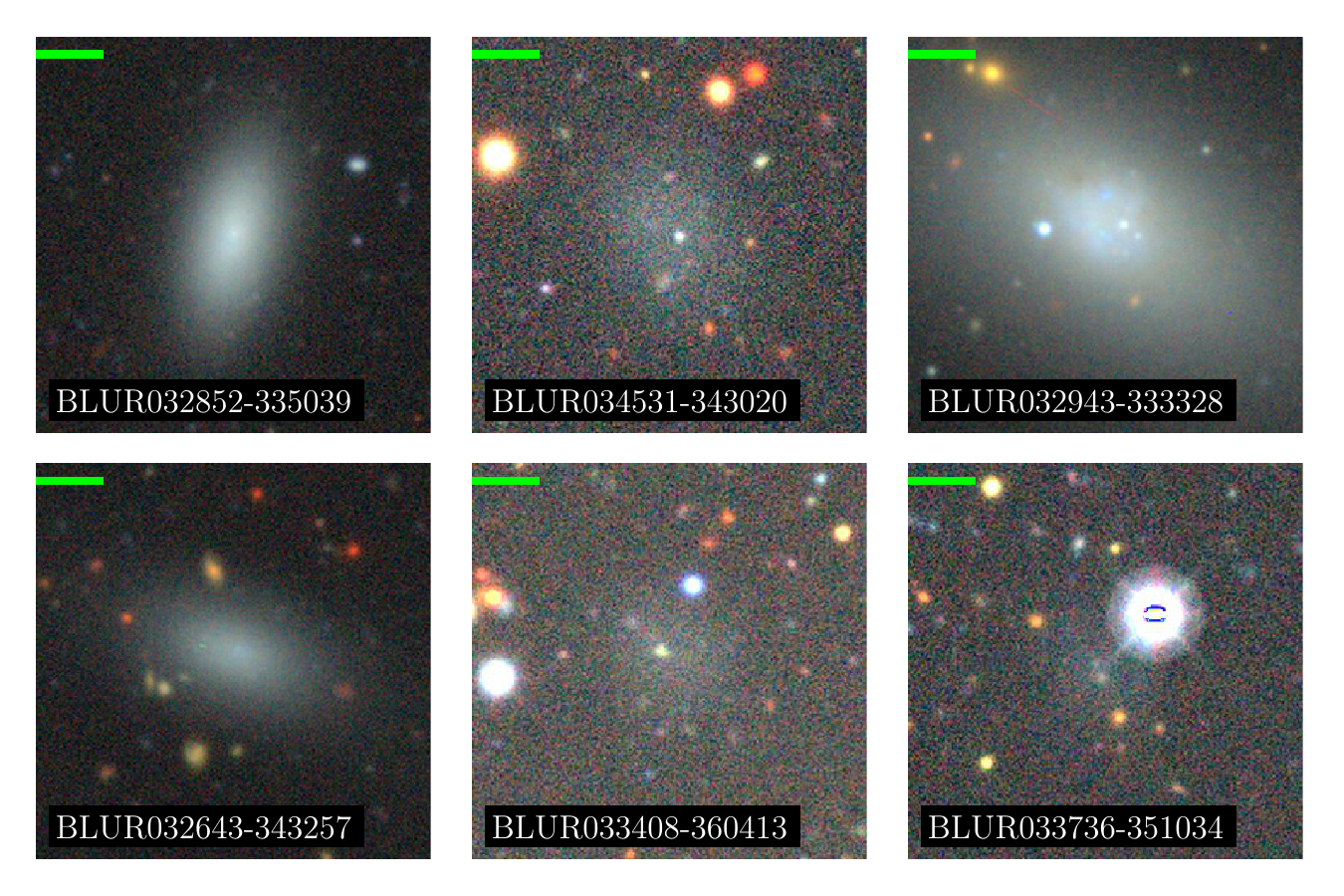}
    \caption{The 6 candidates in our final catalog where we were unable to conclusively determine their nucleated status and were left classified as ``uncertain". This ambiguity primarily arises from the potential NSC structure being too faint or offset from the galaxy center.}
    \label{fig:uncertain_nucl}
\end{figure*}

\section{Blurs Objects Not Recovered By Automated Searches \label{appendix:weFound_NotAutomated}}
In total, we detect 97 objects in our final catalog that were not found by any of the three automated searches of Fornax (V18, V22, and SMUDGes). We highlight the morphologies of a select sample of these objects missed by automated detection in Figure \ref{fig:wefound_notAutomated}. A substantial amount are notably located near bright stars or galaxies.

\begin{figure*}[ht!]
    \centering
    \includegraphics[width=1.0\linewidth]{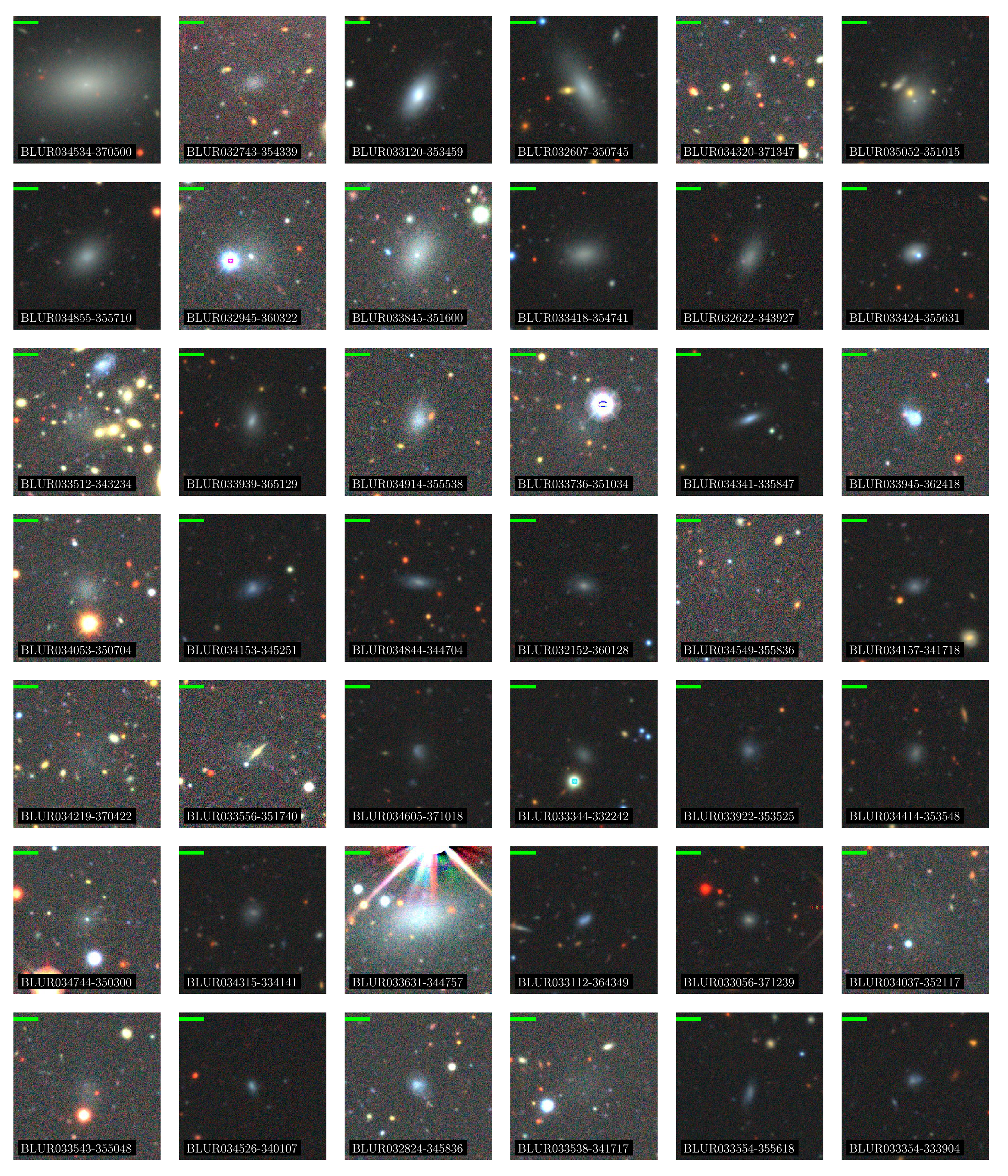}
    \caption{We present DECaLS images of the 42 brightest objects (out of 97 total) we identify in our final catalog that automated searches of Fornax (V18, V22, and SMUDGes) did not detect. These objects are arranged according to their $r$-band apparent magnitude $m_r$, growing fainter in $m_r$ as one moves left to right across each row and from top to bottom. }
    \label{fig:wefound_notAutomated}
\end{figure*}

\section{Objects Not Recovered by Blurs \label{appendix:not_recovered}}
The objects we do not recover from the automated searches of V18, V22, and SMUDGes are shown in Figure \ref{fig:WeMissed_Automated}, while objects not recovered from the visual searches of NGFS, FCC, and P23 are shown in Figure \ref{fig:we-missed-visual}. In general, the objects we do not recover from these catalogs are either a morphology not targeted by our search or too faint in surface brightness to be detected by our methodology.

\begin{figure*}[ht!]
    \centering
    \includegraphics[width=0.9\linewidth]{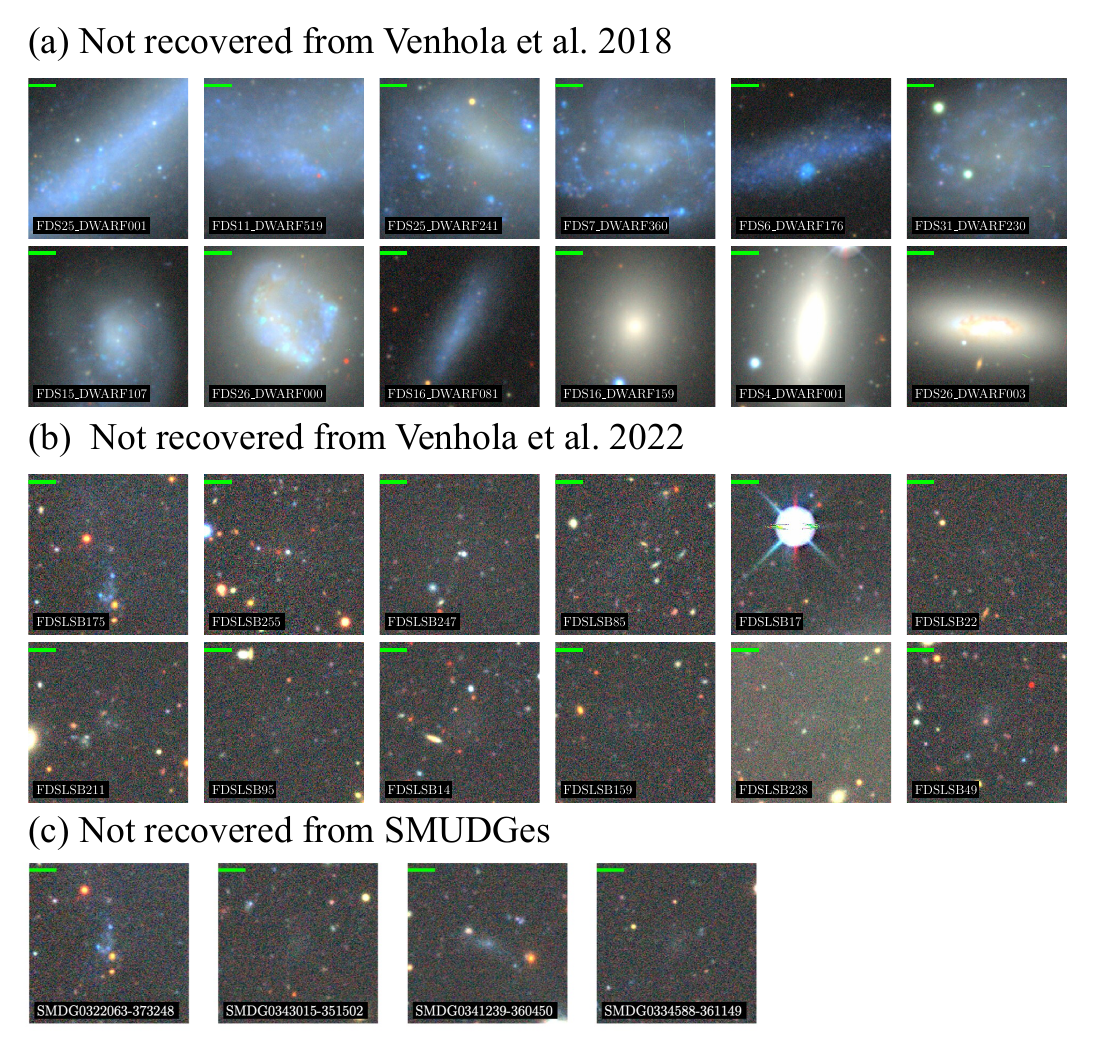}
    \caption{DGs not detected by our Blurs search, but recovered by other programs. (a) DECaLS images of the 12 brightest objects (growing fainter from left to right) in V18 not recovered by Blurs. More than half of the 41 unrecovered objects from V18 are classified as a late-type dwarf, reducing our recovery fraction in V18 for objects with $r_{\mathrm{eff}} \ge 5\arcsec$. (b) Same as the top panel repeated with V22. Most of the objects shown are potentially too faint in surface brightness to be detected visually. (c) The four objects in SMUDGes in the Fornax-overlap region that Blurs did not recover. The left object resembles a late-type galaxy, while the other two objects do resemble proper DG candidates but were possibly missed due to the limit in surface brightness also affecting V22.}
    \label{fig:WeMissed_Automated}
\end{figure*}

\begin{figure*}[ht!]
    \centering
    \includegraphics[width=0.88\linewidth]{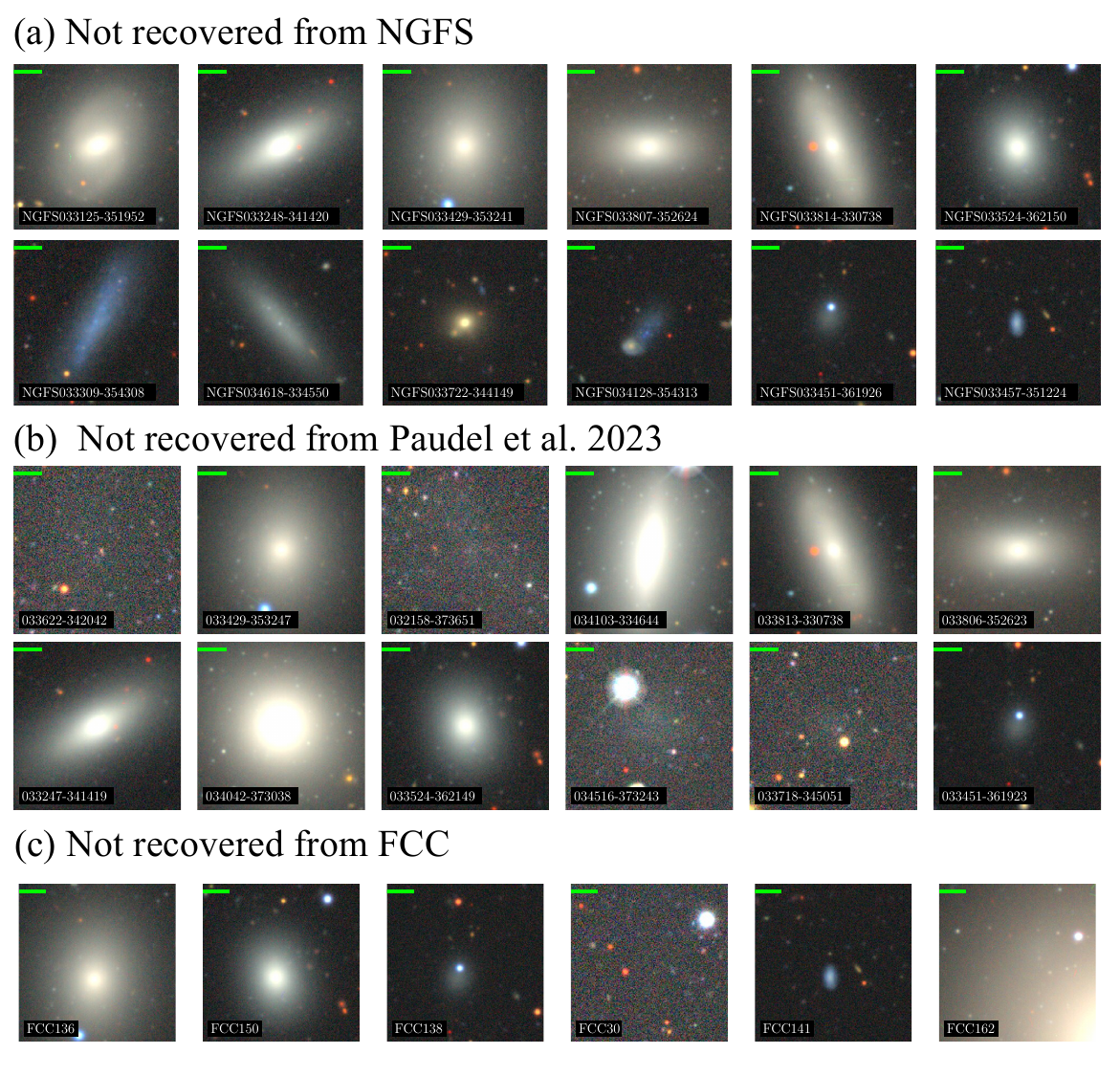}
    \caption{Same as Fig. \ref{fig:WeMissed_Automated} repeated with the visual catalogs discussed in this work: NGFS, FCC, and P23. (a) The 12 brightest DGs (based on $i$-band magnitude) in NGFS not recovered by Blurs. A few objects shown are good DG candidates while the others possess incorrect morphologies for our search. By the end of the bottom row, the galaxy size has significantly shrunk from the first object in the top row, demonstrating that we did not miss many DG candidates over this large range. (b) A sample of 12 objects (out of 22 total) in P23 not recovered by Blurs, ordered by $r_{\mathrm{eff}}$ (instead of brightness), some of which are DGs that may have been too faint for volunteers to find. 
    (c) The six total objects missed from the FCC classified as dE sorted by brightness. Only FCC138 resembles a proper DG candidate from this list.} 
    \label{fig:we-missed-visual}
\end{figure*}

\end{document}